%% file: main.tex
\documentclass{iopjournal}

\usepackage{graphicx}
\usepackage{siunitx}
\usepackage{hyperref}

\usepackage{xcolor}
\usepackage[normalem]{ulem} 

\newif\ifrev
\revfalse     

\usepackage{amsmath, amssymb}
\usepackage{subcaption}

\newcommand{\journalheader}{Plasma Phys. Control. Fusion}
\newcommand{\authorheader}{Orlov et al}
\begin{document}

\articletype{Paper} 

\title{Assessment of 0-D L-H Power Threshold Scaling and Regression Stability in DIII-D with Applied 3D Magnetic Fields}

\author{
  M O Hanson$^{1,2,\dagger}$,
  G R Tynan$^{1,3}$,
  D M Orlov$^{1,\dagger}$
}


\affil{$^1$Center for Energy Research, University of California San Diego, La Jolla, CA 92122, USA}
\affil{$^2$Commonwealth Fusion Systems, Devens, MA 01434, USA}
\affil{$^3$Massachusetts Institute of Technology, Cambridge, MA 02139, USA}

\email{dorlov@ucsd.edu}

\footnote{$\dagger$ M O Hanson and D M Orlov contributed equally to this work.}


\begin{abstract}
A dedicated database of L–H transition discharges has been constructed for the DIII-D tokamak to assess the effect of applied three-dimensional (3D) magnetic fields on the H-mode power threshold. The database spans a broad range of plasma conditions, including discharges with resonant magnetic perturbations (RMPs) applied for edge-localized mode control. Transition times were identified manually and subjected to a series of filtering criteria designed to reduce uncertainties associated with absorbed power, beam modulation, and fast-ion losses. Resonant and non-resonant components of the applied 3D fields were characterized using magnetic equilibrium reconstruction and spectral analysis.

The resulting database is compared against the 2008 ITPA multi-machine L–H power threshold scaling and is used to evaluate the extent to which standard zero-dimensional (0-D) empirical scalings capture the impact of applied 3D fields on H-mode access. Significant scatter and systematic deviations from existing empirical scalings are observed, including for discharges without applied RMPs. Comparisons with TRANSP modeling indicate that commonly used empirical estimates of fast-ion losses may substantially underpredict the actual loss power, particularly at low plasma current and in the presence of applied non-axisymmetric fields.

Regression studies including various 3D field metrics show that while applied perturbations can modify the L–H power threshold, their effects are not robustly captured by simple global scaling parameters alone. 
A comprehensive unconstrained regression incorporating both resonant and non-resonant 3D field metrics yields a substantial residual root-mean-square error (RMSE) of 70\% and produces nonphysical scaling exponents---such as a dramatic inflation of the plasma surface area exponent to 2.79. These results indicate that unresolved hidden-variable dependencies can strongly influence empirical threshold parameterizations even within comparatively restricted single-machine datasets.

Extrapolation of the resulting regressions to ITER-relevant operating conditions yields substantial uncertainty in the predicted threshold power, with confidence intervals spanning a large fraction of the anticipated auxiliary heating capability. These results suggest that hidden machine-specific and edge-localized physics play a significant role in determining the transition threshold and limit the predictive capability of purely 0-D empirical scaling approaches. The implications of these findings for ITER and future reactor-scale devices operating with applied 3D fields are discussed.
\end{abstract}



\input{Section1_Intro}          
\input{Section2_Database}       
\input{Section3_3DMP}      
\input{Section4_Power}       
\input{Section5_Results}       
\input{Section6_Discussion}   
\input{Section7_Conclusions}    

\ack{
We wish to dedicate this work to the memory of our late co-author and mentor, Dr. Todd E. Evans. His foundational contributions at the inception of this project and his profound, enduring mentorship were the primary inspiration for this research; his unexpected passing during its development is deeply felt, and he is missed as both a brilliant colleague and a friend. The authors express their gratitude to Dr. Brian Grierson for his invaluable guidance and collaborative support throughout this work. We also thank Dr. Alexey Arefiev for his vital insights and academic oversight. Technical assistance from Brian Sammuli with database filtering using TokSearch is gratefully acknowledged. We thank the OMFIT development team for providing the integrated modeling environment that enabled this analysis to be performed efficiently. Finally, the authors thank Bill Heidbrink, Mike Van Zeeland, Kathreen Thome, Oak Nelson, and Lothar Schmitz for many fruitful and inspiring discussions on fast-ion transport and L-H transition physics.
}

\funding{%
This material is based upon work supported by the U.S. Department of Energy, Office of Science, Office of Fusion Energy Sciences, using the DIII-D National Fusion Facility, a DOE Office of Science user facility, under Award(s) DE-FC02-04ER54698, DE-SC0026433, and DE-FG02-05ER54809.
}

\roles{
D M Orlov: Conceptualization, Methodology, Supervision, Formal analysis, Writing -- original draft, Writing -- review and editing.\\
M O Hanson: Investigation, Data Curation, Software, Formal analysis, Validation, Visualization, Writing -- original draft, Writing -- review and editing\\
G R Tynan: Supervision, Conceptualization, Writing -- review and editing.
}

\data{Data supporting the findings of this study, including the processed DIII-D L-H transition database parameters, high-resolution EFIT equilibrium reconstructions, vacuum 3D magnetic field spectra from SURFMN, and transport modeling datasets from TRANSP, will be made available from the corresponding author upon reasonable request.}

\suppdata{%
Additional figures and sensitivity scans (including extended rotation and impurity scans) are provided as supplementary material in PDF format.
}

\section*{Conflict of Interest}
The authors declare that they have no conflicts of interest.

\section*{Ethics statement}
This research did not involve any human participants, human data, or animal subjects, and therefore did not require ethical approval.

\section*{References}
\bibliographystyle{plain}

\newpage

Disclaimer: This report was prepared as an account of work sponsored by an agency of the United States Government. Neither the United States Government nor any agency thereof, nor any of their employees, makes any warranty, express or implied, or assumes any legal liability or responsibility for the accuracy, completeness, or usefulness of any information, apparatus, product, or process disclosed, or represents that its use would not infringe privately owned rights. Reference herein to any specific commercial product, process, or service by trade name, trademark, manufacturer, or otherwise does not necessarily constitute or imply its endorsement, recommendation, or favoring by the United States Government or any agency thereof. The views and opinions of authors expressed herein do not necessarily state or reflect those of the United States Government or any agency thereof.


\end{document}

%% file: Section1_Intro.tex
\section{Introduction}

Access to the high-confinement mode (H-mode) is essential for ITER, SPARC \cite{creely2020}, ARC \cite{howard2026_arc}, and future reactor-scale tokamaks to achieve the confinement required for high fusion gain \cite{wagner1982,burrell1987}. At the same time, operation in H-mode introduces edge-localized modes (ELMs), which can produce transient heat loads large enough to damage plasma-facing components in reactor-scale devices \cite{Kuang2020SPARC, Eich2026ARC}. Resonant magnetic perturbations (RMPs) remain one of the leading approaches for ELM suppression and mitigation and are planned for use in ITER through the application of non-axisymmetric magnetic fields generated by internal control coils \cite{evans2013_iter}. Since ITER will likely apply these perturbations prior to the transition into H-mode to avoid the first large ELM, understanding the effect of applied three-dimensional (3D) magnetic fields on the L--H power threshold is important for reliable startup scenario development.

The physics of the L--H transition is known to depend on a broad range of plasma and divertor parameters, including density, magnetic field, plasma shape, ion $\nabla B$ drift direction, edge flows, divertor geometry, wall conditions, and heating method. Extensive single-device parameter variations have demonstrated that explicit dependencies on edge safety factor and turbulence structures \cite{yan2019_pop}, divertor configuration geometry \cite{andrew2004}, and ITER-relevant plasma shaping boundaries \cite{gohil2011} introduce critical non-local modifications to the baseline threshold. This behavior is heavily coupled to macro-scale operating parameters, manifesting as distinct variations within separate operational density regimes, such as the spontaneous edge threshold characteristics observed on DIII-D \cite{carlstrom1996} and the non-monotonic low-density branch scaling behaviors systematically documented on Alcator C-Mod \cite{ma2012} and ASDEX Upgrade \cite{ryter2013}. Previous experimental studies have additionally shown that applied non-axisymmetric magnetic perturbations can modify the radial electric field and edge flow shear near the separatrix, potentially increasing the H-mode power threshold \cite{gohil2011}. In DIII-D, sufficiently strong resonant $n=3$ perturbations have been shown to reduce or even locally reverse the edge radial electric field well, thereby weakening the $E\times B$ shear thought to contribute to turbulence suppression at the transition \cite{schmitz2019}. These observations are consistent with previous experimental studies in DIII-D demonstrating modification of the edge radial electric field and edge transport by applied resonant magnetic perturbations \cite{gohil2011, schmitz2019}.

Despite extensive experimental and theoretical work, predictive modeling of the L--H  transition remains difficult. The transition is widely understood to be triggered by localized, turbulence-driven flows at the plasma edge via a Reynolds stress drive \cite{Yan2014}, a microscopic dynamic that cannot be easily parameterized by global 0-D quantities. Consequently, empirical scaling approaches continue to play a central role in reactor projections. The most widely used scaling for the L--H power threshold was developed by Martin et al.\ using the International Tokamak Physics Activity (ITPA) multi-machine database \cite{martin2008itpa}.
Restricting the analysis to ITER-relevant discharges, the resulting regression produced a scaling of the threshold power in terms of the line-averaged density, toroidal magnetic field, and plasma surface area. This scaling remains widely used for ITER scenario planning.

While the ITPA scaling provides a valuable empirical framework, several limitations motivate a dedicated re-examination using modern DIII-D data. First, the original DIII-D contribution to the database was limited to discharges from the early 1990s, prior to many upgrades in diagnostics and plasma control capabilities. Second, the multi-machine nature of the database introduces unavoidable inconsistencies in power accounting, wall conditions, fast-ion loss treatment, and equilibrium reconstruction methodologies. Third, the original scaling does not explicitly include the effect of externally applied 3D magnetic fields, despite their importance for ITER-relevant operating scenarios.

In addition, several studies have suggested that uncertainties associated with the definition of the loss power, particularly fast-ion losses and radiated power corrections, may contribute significantly to the scatter observed in empirical threshold scalings. In DIII-D, fast-ion losses are commonly estimated using simplified empirical models that depend primarily on plasma current, even though neutral beam losses are known to depend strongly on plasma conditions and can increase substantially in the presence of applied 3D fields.

Motivated by these issues, a dedicated database of DIII-D L--H transitions has been constructed to investigate the effect of applied 3D magnetic fields on the H-mode power threshold. The database includes discharges spanning a broad range of plasma conditions, including transitions with resonant and non-resonant perturbations generated by DIII-D I- and error field correction coils. Transition times were manually identified and subjected to a series of filtering criteria designed to reduce uncertainties associated with absorbed power, beam modulation, and fast-ion loss estimates.

The goals of this work are threefold. First, the database is used to assess the extent to which the standard 0--D ITPA scaling captures modern DIII-D L--H transition behavior. Second, the study evaluates whether simple global metrics of applied 3D magnetic perturbations can robustly parameterize the effect of RMPs on the transition threshold. This analysis is specifically motivated by the fact that both resonant and non-resonant magnetic perturbations impact $P_{\text{LH}}$ by directly modifying local turbulence-flow interactions and edge shear layers \cite{Kriete2020}---a localized physical mechanism that global, low-dimensional empirical variables cannot easily map.
Third, the work examines the role of power accounting uncertainties, particularly fast-ion losses, in determining empirical threshold scalings and their extrapolation to ITER-relevant conditions.

The paper is organized as follows. Section 2 describes the construction of the DIII-D L--H transition database and the transition selection methodology. Section 3 discusses the characterization of resonant and non-resonant magnetic perturbations. Section 4 describes the power accounting methodology and the treatment of fast-ion losses. Section 5 presents comparisons with the Martin scaling and examines the dependence of the threshold power on applied 3D fields. Finally, Section 6 discusses the implications of these results for empirical scaling approaches and reactor extrapolation.

%% file: Section2_Database.tex
\section{Database Construction and Transition Selection}
\label{sec:Database}

A dedicated database of DIII-D L--H transitions was constructed to investigate the effect of applied three-dimensional (3D) magnetic perturbations on the H-mode power threshold. Unlike the original DIII-D contribution to the ITPA L--H threshold database, which primarily consisted of discharges from the early 1990s, the present database incorporates more recent DIII-D experiments with improved diagnostic coverage, equilibrium reconstruction capability, and magnetic perturbation control. The resulting dataset spans a broad range of plasma conditions, including discharges with resonant magnetic perturbations (RMPs), non-resonant magnetic perturbations (NRMPs), and nominally axisymmetric configurations.

The database was intentionally restricted to ITER-relevant operating conditions following the general philosophy of the 2008 ITPA study \cite{martin2008itpa}. Only deuterium single-null discharges with favorable ion $\nabla B$ drift direction were retained for the primary analysis. To reduce variations associated with known machine configuration dependencies and improve consistency with reactor-relevant operating scenarios, a rigorous multi-step conditioning filter was applied to the initial data set \cite{orlov2014}. Within this database conditioning workflow, restricting the plasma configuration to single-null divertors removed 9 discharges, while isolating purely favorable ion $\nabla B$ drift configurations eliminated 15 discharges. The enforcement of a high-density branch threshold constraint ($n_{e20} > 0.2 \times 10^{20}$~m$^{-3}$) removed 75 discharges, non-optimal plasma-wall clearances ($\text{gap}_{\text{in}}/\text{gap}_{\text{out}} < 0.05$~m) removed 17 discharges, and the filtering of purely ohmic or electron cyclotron heating (ECH) only scenarios excluded 1 and 19 discharges, respectively. Applying these strict criteria successfully isolated a core conditioned subset of 60 highly comparable transitions for primary regression analysis.

Candidate discharges were first identified using a combination of SQL-based shot filtering and server-side data processing implemented through TokSearch \cite{sammuli2018} within the OMFIT framework \cite{meneghini2015}. Initial filtering selected discharges with plasma current duration exceeding approximately 1~s and toroidal magnetic field strength greater than 1~T. This broad preselection was then refined by evaluating derived magnetic perturbation signals associated with the DIII-D I-coils and correction coils.

To identify discharges with potential applied 3D fields, harmonic magnetic field amplitudes generated by the I-coils and C-coils were evaluated using derived signals corresponding to the normal magnetic field perturbation at the plasma boundary. Thresholds were applied to distinguish intentional perturbations from background noise. Discharges with perturbation amplitudes below approximately 1~G were treated as nominally axisymmetric, while larger amplitudes were classified as candidate RMP or NRMP cases depending on the perturbation structure.

L--H transition times were identified manually using a combination of divertor D$_\alpha$ emission signals and line-averaged electron density measurements. The transition was identified by the characteristic rapid decrease in divertor recycling emission accompanied by a simultaneous rise in plasma density. Manual identification was chosen over fully automated methods in order to maintain consistency across operating scenarios and to reduce false-positive identifications associated with transient confinement changes or beam modulation events.

Several additional filtering steps were introduced to reduce uncertainty in the inferred threshold power. First, strongly overdriven H-mode transitions were removed. In many DIII-D experiments, heating power is rapidly increased well above the transition threshold in order to guarantee H-mode access. In such cases, the actual threshold power becomes poorly constrained because the transition may occur during the slowing-down time of injected neutral beam ions. To reduce ambiguity in the absorbed power term, discharges containing large heating power steps within approximately 200~ms of the transition were excluded.

Second, discharges with strong neutral beam power modulation near the transition were removed. Large-amplitude beam modulation introduces significant uncertainty in both the absorbed power and the time derivative of the stored plasma energy. Since the transition may occur during either the rising or falling portion of the modulation cycle, the corresponding threshold power can become poorly constrained.

Third, discharges with significant counter-current neutral beam injection were flagged separately. Previous studies have shown that fast-ion losses can increase substantially in the presence of counter-injected beams, particularly at lower plasma current. Since simplified fast-ion loss models are less reliable in these conditions, these discharges were treated with reduced confidence in subsequent analysis.

Following transition identification and filtering, each database entry was assigned a confidence classification reflecting the estimated uncertainty in the loss power calculation. The confidence ranking incorporated factors including beam modulation amplitude, proximity to heating power ramps, uncertainty in fast-ion losses, and overall signal quality. This procedure was designed to preserve a larger exploratory dataset while allowing higher-confidence subsets to be isolated for regression studies.

To preserve statistical coverage while retaining the ability to isolate higher-quality transitions, the database was separated into multiple confidence categories. Higher-confidence transitions corresponded to discharges with relatively steady heating power, minimal beam modulation, reduced ambiguity in transition timing, and more reliable estimates of absorbed power and fast-ion losses. Lower-confidence transitions were retained primarily for exploratory analysis and trend identification but were excluded from selected regression studies requiring tighter control of systematic uncertainties. Of the 192 total database entries, 99 transitions were classified with a highest confidence rating of 1, 32 transitions achieved a confident rating of 2, and 61 entries were assigned a rating of 3 due to mild neutral beam variance or edge scaling fluctuations.

The final database contains 192 DIII-D L--H transitions spanning a range of plasma densities, magnetic fields, plasma currents, heating powers, and applied magnetic perturbation configurations. Both resonant and non-resonant perturbation structures are represented, including discharges with dominant $n=1$, $n=2$, and $n=3$ toroidal mode components generated by the DIII-D I-coil and correction coil systems. The database additionally includes a subset of nominally axisymmetric transitions used for comparison against standard empirical threshold scalings. A summary of the principal database characteristics and operating parameter ranges is provided in table~\ref{tab:database_summary}.

\begin{table}[t]
\centering
\caption{Summary of the DIII-D L–H transition database used in the present study. The database spans a broad range of ITER-relevant operating conditions and includes both resonant and non-resonant magnetic perturbation configurations. For archival tracking and exact reproducibility, the complete list of the 192 specific DIII-D shot numbers, along with their manually identified transition time slices, is tabulated and provided in the electronic supplementary material.}
\label{tab:database_summary}
\begin{tabular}{lc}
\hline
\textbf{Metric} & \textbf{Value} \\
\hline
Shots fitting 2008 ITPA criteria & 60 \\
Upper single-null discharges & 12 \\
Lower single-null discharges & 171 \\
Double-null discharges & 9 \\
Favorable ion $\nabla B$ drift & 172 \\
Deuterium plasmas & 181 \\
Hydrogen plasmas & 6 \\
Helium plasmas & 5 \\
Neutral beam heating power & 0--6~MW \\
Electron cyclotron heating power & 0--3~MW \\
Plasma current & 0.5--1.6~MA \\
Toroidal magnetic field & 1--2.1~T \\
$dR_{\mathrm{sep}}$ & $-0.40$--0.06~m \\
Plasma surface area & 49--55~m$^2$ \\
L-mode line-averaged electron density & $(1.2$--$5.1)\times10^{19}$~m$^{-3}$ \\
$q_{95}$ & 3--9 \\
Elongation $\kappa$ & 1.67--1.93 \\
Inner gap & 0.02--0.18~m \\
Outer gap & 0.05--0.14~m \\
\hline
\end{tabular}
\end{table}



Compared to the original ITPA multi-machine database, the present single-machine approach sacrifices parameter-space breadth in exchange for improved consistency in diagnostic methodology, equilibrium reconstruction, wall conditions, plasma control, and power accounting. While this restriction limits the accessible extrapolation range, it substantially reduces hidden machine-to-machine variability that may otherwise obscure comparatively subtle physical trends. Single-device studies additionally allow operational details such as divertor geometry, beam modulation behavior, fast-ion confinement assumptions, and perturbation coil configurations to be treated more consistently than is generally possible within large international databases. This improved internal consistency is particularly important when attempting to isolate the comparatively small threshold modifications associated with applied 3D magnetic perturbations. It must be explicitly acknowledged, however, that DIII-D utilizes a graphite/carbon first wall. Cross-machine comparisons have demonstrated that carbon-wall configurations exhibit a systematically higher baseline threshold power than devices operating with metallic plasma-facing components, such as tungsten or beryllium/tungsten ($P_{\text{LH,W}} < P_{\text{LH,C}}$) \cite{ryter2013survey, Maggi2014}. While this systematic offset shifts the absolute power levels required for direct reactor extrapolation, this uniform carbon-wall dataset remains fully appropriate for isolating and assessing the relative structural impacts of 3D non-axisymmetric magnetic perturbations.

%% file: Section3_3DMP.tex
\section{Characterization of Three-Dimensional Magnetic Perturbations}
\label{sec:3dfields}

To quantify the effect of applied non-axisymmetric magnetic fields on the L--H power threshold, the resonant and non-resonant components of the magnetic perturbation spectrum were evaluated for each discharge in the database. In DIII-D, externally applied 3D fields are generated primarily by two coil systems: the internal I-coils used for edge-localized mode (ELM) suppression—which consist of six upper and six lower rows of window-pane coils mounted on the radially outboard side of the vacuum vessel that produce a normalized perturbation of $\delta B / B_T \approx 10^{-4}$ \cite{Jackson2003,evans2004_prl}—and the external correction (C) coils used for error field correction.
Depending on the applied phasing and operating configuration, these coils can generate perturbations with dominant toroidal mode numbers $n=1$, $n=2$, or $n=3$.

The equilibrium magnetic geometry for each discharge was reconstructed using EFIT \cite{lao1985}. The reconstructed equilibria were then used as input to the SURFMN code \cite{schaffer2008}, which evaluates the resonant magnetic perturbation spectrum in straight-field-line coordinates using Fourier decomposition of the perturbed equilibrium fields. This procedure allows the resonant magnetic field strength to be evaluated as a function of toroidal mode number $n$, poloidal mode number $m$, and normalized poloidal flux $\Psi_N$.

For each toroidal mode, the resonant component of the perturbation was identified along the pitch-resonance condition defined by

\begin{equation}
q = \frac{m}{n},
\end{equation}

where $q$ is the safety factor and $m$ and $n$ are the poloidal and toroidal mode numbers, respectively. The resonant magnetic perturbation magnitude was quantified using the radial magnetic field component $\delta B_r$ evaluated at selected flux surfaces near the plasma edge.

Several flux surface ranges were examined in order to characterize the radial structure of the perturbation spectrum near the separatrix. For each discharge, the resonant fields were evaluated and recorded at three specific edge radial coordinates: $\Psi_N = 0.95$, $0.97$, and $0.99$, with range windows spanning $\pm 0.01$ around each target value to smooth out highly localized radial alignment variations. Across the unreduced database, spectral analysis identified a total of 455 resonant surfaces within the $0.94$--$0.9599$ flux interval, 803 surfaces within the $0.96$--$0.9799$ interval, and 1339 surfaces within the outermost $0.98$--$0.999$ boundary range for fields with relative magnitudes greater than $1\times 10^{-8}$. When filtering for fields relevant to threshold modification by omitting faint baseline noise below $1\times 10^{-4}$, a total of approximately 400 resonant surfaces remain, heavily concentrated from $\Psi_N = 0.98$ outwards.

The perturbation strength was parameterized using the normalized quantity

\begin{equation}
\frac{\delta B_r}{B_T},
\end{equation}

where $B_T$ is the toroidal magnetic field evaluated on axis. Normalizing the perturbation amplitude by the equilibrium toroidal field allows comparison across discharges with different operating magnetic fields and facilitates comparisons with previous RMP studies performed on DIII-D and other tokamaks.

In addition to resonant perturbations, non-resonant magnetic field components were also evaluated. While resonant perturbations are expected to produce the strongest coupling to rational magnetic surfaces, non-resonant components can still influence edge transport through modifications to plasma response, flow damping, magnetic topology, and stochastic layer formation. The relative contributions of resonant and non-resonant perturbations were therefore retained separately in the database analysis.

\begin{figure}[t]
    \centering
    \includegraphics[width=0.48\textwidth]{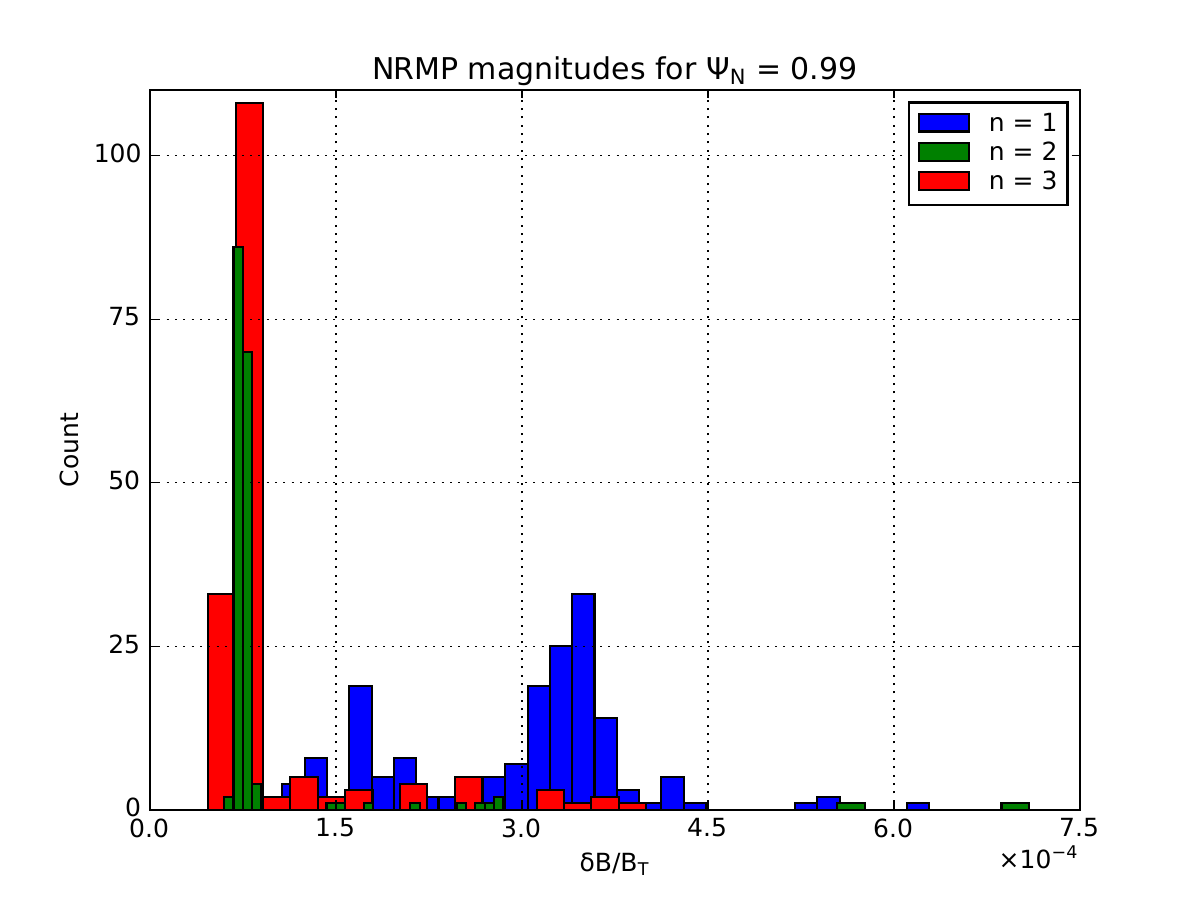}
    \includegraphics[width=0.48\textwidth]{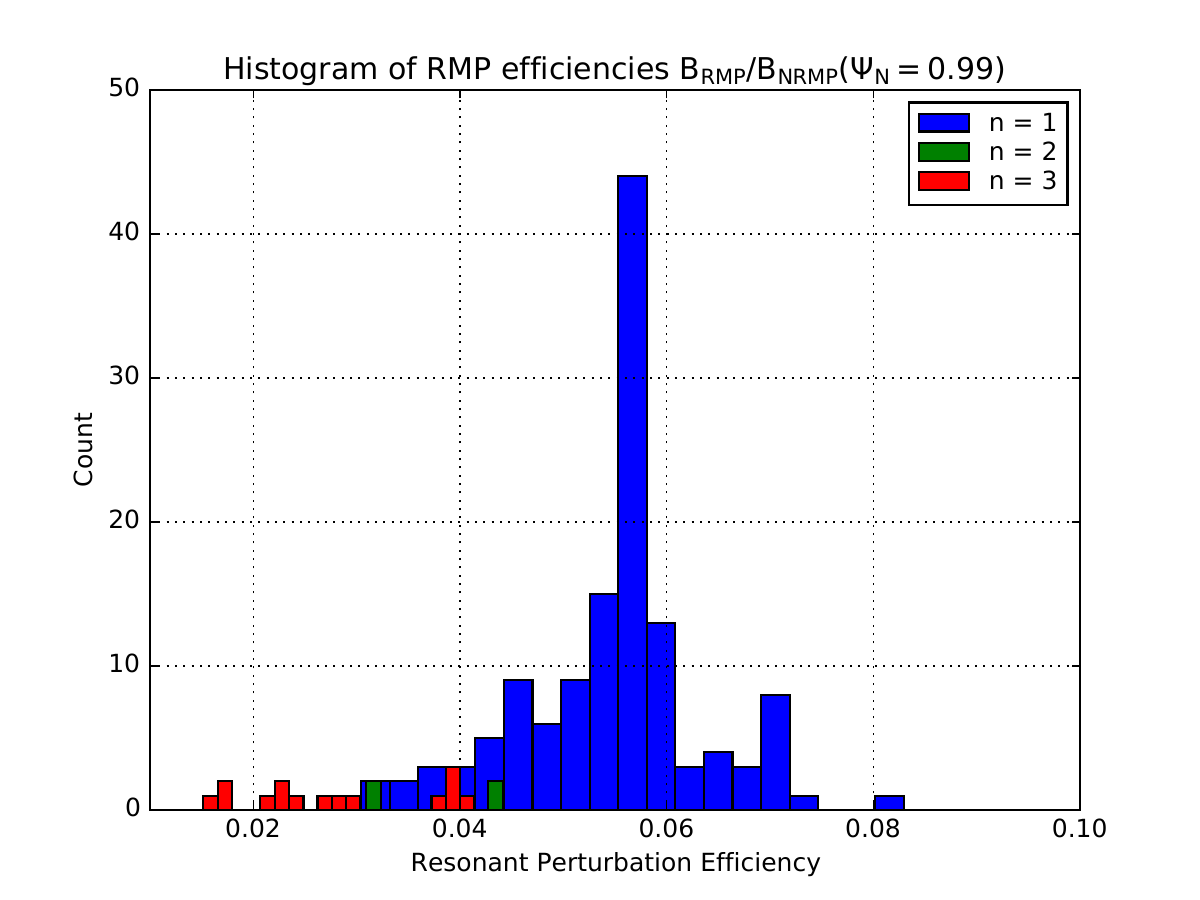}
    \caption{Characterization of the applied three-dimensional magnetic perturbation spectrum in the DIII-D L--H transition database. 
    (a) Distribution of relative magnitudes for the non-resonant components of the applied 3D fields. 
    (b) Magnetic perturbation efficiency for the dominant $n=1$, $n=2$, and $n=3$ toroidal mode components evaluated near the plasma edge at $\Psi_N=0.99$. 
    The perturbation efficiency is defined as the ratio of the resonant radial magnetic field component to the integrated non-resonant contribution at the same radial location \cite{schaffer2008}.}
    \label{fig:perturbation_efficiency}
\end{figure}

The resulting perturbation spectrum exhibits substantial variability across the database. Discharges with applied ELM suppression fields typically contained normalized perturbation magnitudes on the order of $10^{-4}$ to $10^{-3}$ near the plasma edge, with the strongest resonant coupling generally associated with applied $n=3$ perturbations. In contrast, nominally axisymmetric discharges and cases with only weak correction fields exhibited significantly smaller resonant amplitudes. As shown by the non-resonant database demographics in Figure~\ref{fig:perturbation_efficiency}(a), the $n=2$ and $n=3$ non-resonant components are mostly small, with the majority of the population concentrated around $7.5\times 10^{-5}$, whereas the $n=1$ non-resonant components exhibit a broader distribution with an average relative magnitude of $3\times 10^{-4}$. Furthermore, the calculated vacuum magnetic perturbation efficiency evaluated near the plasma edge at $\Psi_N = 0.99$ and plotted in Figure~\ref{fig:perturbation_efficiency}(b) shows that the $n=1$ mode peaks strongly around $5\%$ (indicating that $95\%$ of the field at this radial coordinate is non-resonant), while the $n=3$ component demonstrates a lower average efficiency of approximately $3\%$.

Previous DIII-D studies have shown that sufficiently strong resonant perturbations can reduce or locally reverse the edge radial electric field well, thereby weakening the $E\times B$ shear associated with turbulence suppression at the L--H transition. Motivated by these observations, several regression metrics based on resonant field amplitude, toroidal mode number, and radial location were explored in the present study to determine whether applied 3D fields can be incorporated into empirical threshold scaling relations.

The present study focuses primarily on experimentally accessible vacuum-field perturbation metrics derived from equilibrium reconstruction and spectral decomposition. More sophisticated plasma-response calculations using resistive MHD models were beyond the scope of the present work. Consequently, the perturbation quantities used here should be interpreted primarily as proxies for the effective edge perturbation rather than direct measurements of the fully self-consistent plasma response.

Although global perturbation metrics provide a convenient parameterization for database studies, it is important to note that they do not fully capture the plasma response to externally applied fields. Screening, resonant field amplification, edge stochasticity, and rotational shielding are all known to depend sensitively on local equilibrium and kinetic conditions. Consequently, the perturbation metrics used here should be interpreted primarily as experimentally accessible proxies for the underlying edge physics rather than complete descriptors of the plasma response.

%% file: Section4_Power.tex
\section{Power Accounting and Fast-Ion Loss Treatment}
\label{sec:power}

The power crossing the separatrix at the time of the L--H transition was quantified using a modified form of the standard ITPA loss power definition. In the original 2008 ITPA scaling study, the loss power was defined as

\begin{equation}
P_{\mathrm{Loss}} = P_{\mathrm{Ohm}} + P_{\mathrm{Abs}} - \frac{dW}{dt} - P_{\mathrm{F,loss}},
\label{eq:ploss_itpa}
\end{equation}

where $P_{\mathrm{Ohm}}$ is the ohmic heating power, $P_{\mathrm{Abs}}$ is the absorbed auxiliary heating power, $dW/dt$ is the time derivative of the plasma stored energy determined from MHD equilibrium reconstruction, and $P_{\mathrm{F,loss}}$ represents fast-ion and prompt orbit losses \cite{martin2008itpa}. In the present study, an additional correction was included for core radiated power,

\begin{equation}
P_{\mathrm{Loss}} = P_{\mathrm{Ohm}} + P_{\mathrm{Abs}} - \frac{dW}{dt} - P_{\mathrm{F,loss}} - P_{\mathrm{Rad}}^{\mathrm{core}},
\label{eq:ploss_modified}
\end{equation}

where $P_{\mathrm{Rad}}^{\mathrm{core}}$ was dynamically estimated using DIII-D bolometry measurements \cite{leonard1995}.

The absorbed auxiliary heating power was calculated from the combination of neutral beam injection (NBI) and electron cyclotron heating (ECH), accounting for beam shine-through where available. While $P_{\text{Abs}}$ is treated as a single global scalar within this 0-D framework, it must be emphasized that the specific allocation of power into the ion versus electron channels plays a critical role in transition efficacy. This dependency is particularly acute at lower plasma densities where collisional thermal equilibration between species is poor. Under weak coupling conditions, direct ion heating via NBI is substantially more effective at driving the edge ion pressure gradient and triggering the L--H transition, whereas electron-dominated heating via ECH typically results in a significantly elevated total power threshold \cite{ryter2013survey}. Treating these fundamentally distinct heating channels as a homogeneous sum introduces a hidden systematic variance into empirical regressions.

The inclusion of core radiated power is motivated by the expectation that the L--H transition is governed primarily by the local power flowing across the separatrix into the edge and scrape-off-layer regions rather than the total externally injected power. Previous multi-machine scaling studies often omitted this term because reliable bolometric measurements were not uniformly available across all contributing devices \cite{martin2008itpa}. Restricting the present analysis to DIII-D enables a more consistent and localized treatment of core radiated power losses.

The absorbed auxiliary heating power was calculated from the combination of neutral beam injection (NBI) and electron cyclotron heating (ECH), accounting for beam shine-through where available. Particular care was taken to exclude discharges with large beam power ramps or strong beam modulation near the transition time, since these conditions introduce severe ambiguity in the effective absorbed power. Because the heating power injected via neutral beams is not absorbed instantaneously by the background plasma, its transfer is mediated by finite slowing-down timescales. Profiles of these slowing-down times calculated across the core plasma region ($\Psi_N \sim 0.5$) just before the transition yield characteristic times of $60\text{--}100\text{~ms}$. To accurately simulate this behavior on a database scale, a baseline $100\text{~ms}$ Gaussian causal smoothing window was formally applied to the raw NBI power signals.

The treatment of fast-ion losses represents one of the largest systematic uncertainties in empirical L--H threshold scaling studies. In the original ITPA database, DIII-D fast-ion losses were estimated using a simplified, empirical algebraic expression depending entirely on the total plasma current $I_P$ (in Amperes) and the absorbed power \cite{martin2008itpa}:

\begin{equation}
P_{\mathrm{F,loss}} = P_{\mathrm{Abs}} \times \exp\left(\frac{3.3 - I_P/10^6}{100}\right).
\label{eq:fast_ion_empirical}
\end{equation}

While computationally convenient for large multi-machine database operations, this zero-dimensional approach neglects critical physical dependencies associated with explicit beam-injection geometries, local plasma equilibrium configurations, collisionality profiles, loss-orbit topologies, and non-axisymmetric magnetic perturbations \cite{vanzeeland2015}.

\begin{figure}[t]
    \centering
    \includegraphics[width=0.75\textwidth]{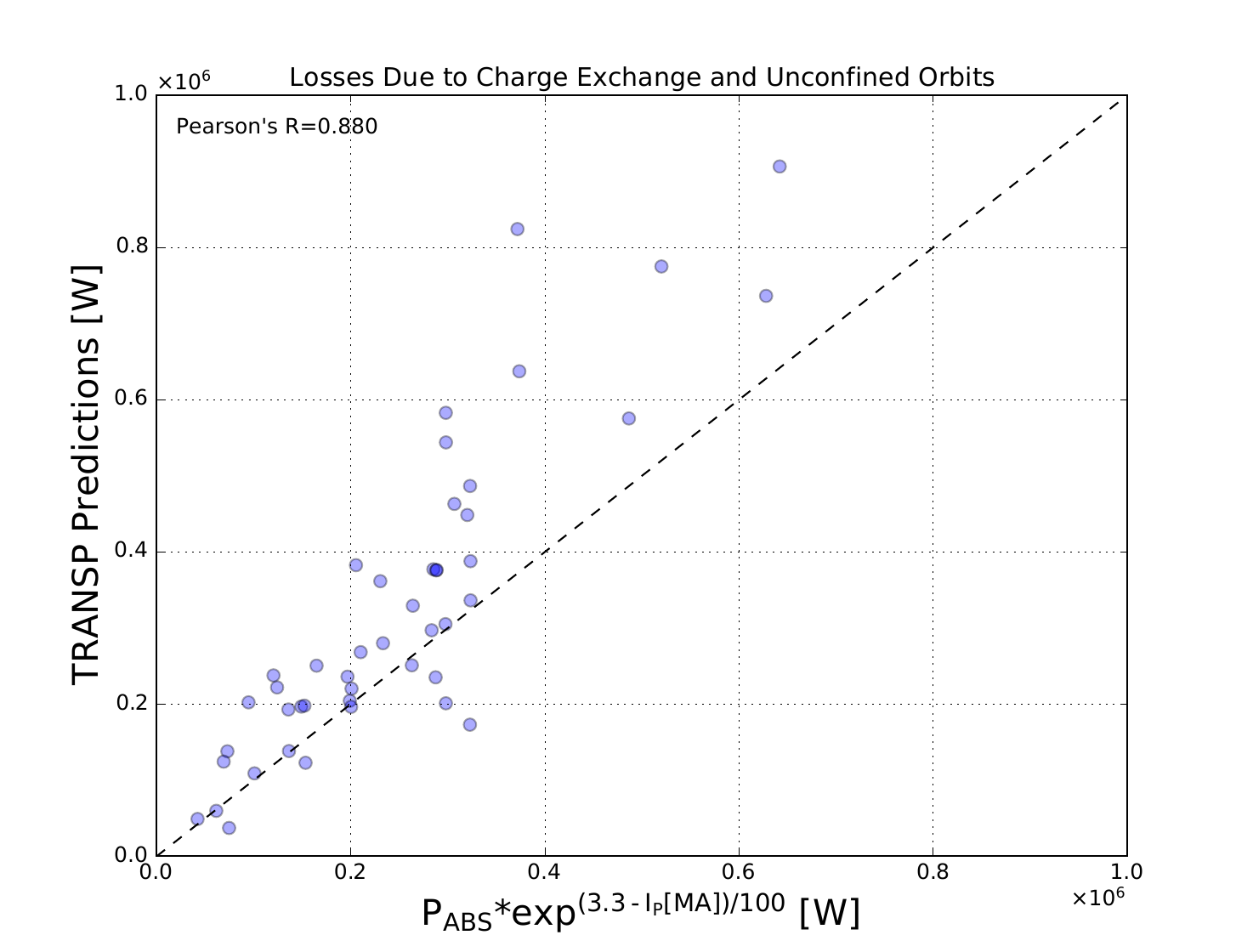}
    \caption{Comparison between the empirical fast-ion loss model used in the original DIII-D contribution to the ITPA database and TRANSP-derived fast-ion loss estimates for the 45 successfully converged discharges within the present database. The black dashed line indicates a perfect 1:1 correlation. Agreement is reasonable for small loss powers ($<0.3\text{~MW}$), while the empirical model systematically underpredicts losses as the plasma current decreases.}
    \label{fig:transp_empirical_fastion}
\end{figure}

To assess the validity of this empirical approximation, transport simulations were systematically attempted for the entire database using the 1.5D time-dependent transport code TRANSP \cite{pankin2025}. Out of the 192 total database entries, 45 discharges successfully converged to a rigorous solution; tracking failures in the remaining runs stemmed primarily from a lack of high-fidelity edge charge-exchange recombination (CER) datasets or quality impurity density profile fits. TRANSP calculations were utilized to solve for fast-ion and prompt orbit losses under experimentally reconstructed plasma parameters. Comparisons between these TRANSP-derived losses and the simplified empirical model from equation~\eqref{eq:fast_ion_empirical} revealed reasonable quantitative agreement only within a limited subset of high-current discharges characterized by comparatively small loss fractions. The overall magnitude of the resulting fast-ion correction is detailed in the scatter comparison in figure~\ref{fig:transp_empirical_fastion}.

\begin{figure}[t]
    \centering
    \includegraphics[width=0.75\textwidth]{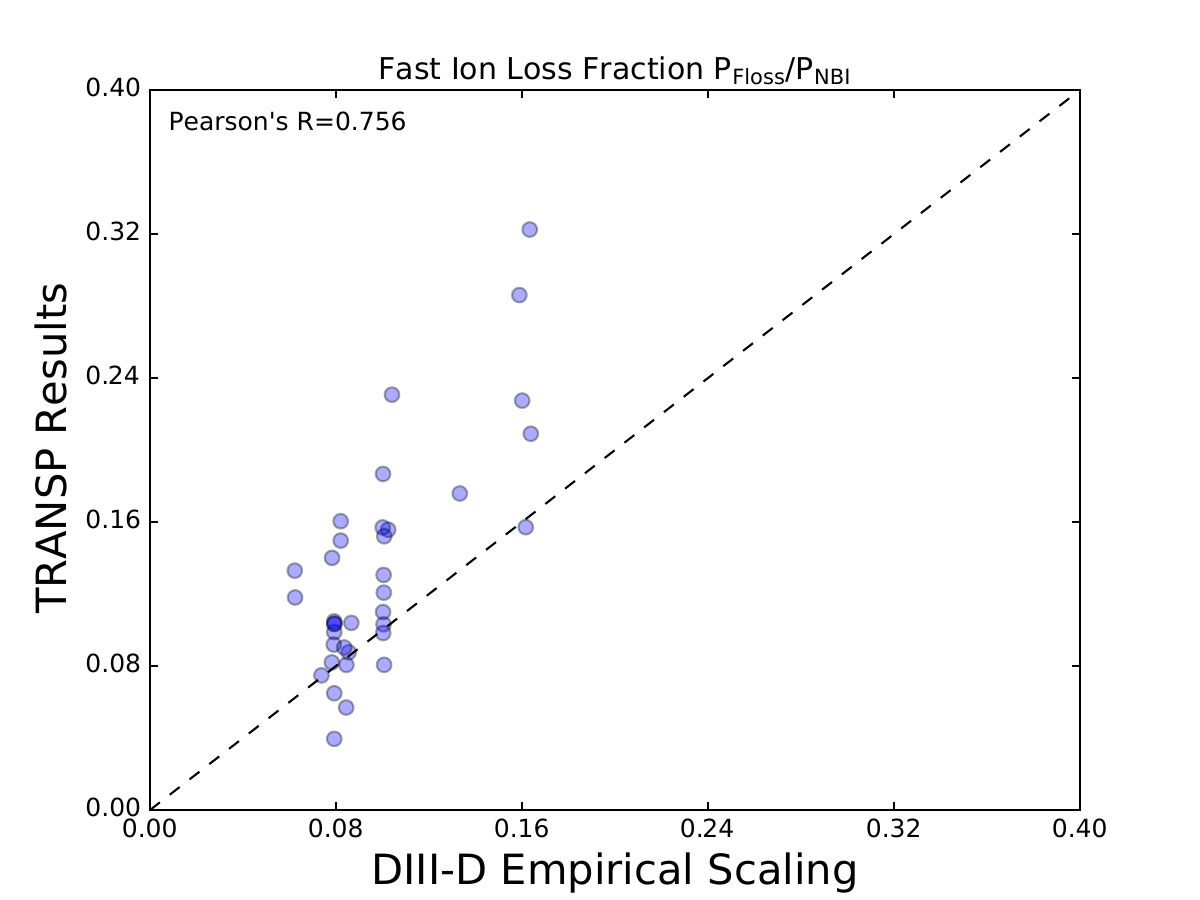}
    \caption{Fast-ion loss power normalized to injected neutral beam power ($P_{\mathrm{Floss}}/P_{\mathrm{NBI}}$) plotted as a function of the underlying DIII-D empirical scaling metrics. Most cases exhibit losses spanning at least several percent of the injected beam power, with multiple TRANSP predictions exceeding 30\%. Note that these calculations do not include additional fast-ion transport driven by applied 3D magnetic fields.}
    \label{fig:fastion_loss_fraction}
    \end{figure}

At lower plasma current, the empirical model systematically underpredicted fast-ion losses relative to TRANSP calculations. As illustrated in the fast-ion loss fraction distribution in figure~\ref{fig:fastion_loss_fraction}, several low-current discharges exhibit actual losses exceeding 30\% of the total injected neutral beam power, whereas the empirical model limits its prediction to approximately 16\%. Furthermore, previous studies have shown that externally applied resonant magnetic perturbations can substantially enhance prompt fast-ion losses and neutral beam orbit losses through stochastic orbit transport and modified magnetic topology. Detailed transport modeling and phase-space mapping of energetic particles confirm that the deliberate introduction of non-axisymmetric 3D fields disrupts the constants of motion governing drift-orbit trajectories, driving rapid prompt orbit losses across the separatrix boundary \cite{vanzeeland2015}. Because the TRANSP simulations executed here do not formally incorporate these explicit 3D magnetic perturbation fields, the calculated fractions must be interpreted as a conservative lower bound of the actual fast-ion loss term.

These effects are particularly important because the fast-ion loss term enters directly into the calculation of the power crossing the separatrix. Underestimating fast-ion losses artificially inflates the calculated value of $P_{\mathrm{Loss}}$, which subsequently injects a systematic upward bias into empirical threshold regressions. Since the true magnitude of the fast-ion correction can approach a substantial fraction of the total injected neutral beam power, inaccuracies in this term contribute heavily to the large residual scatter observed in existing empirical scaling relations.

Despite these limitations, performing fully self-consistent, multi-dimensional fast-ion transport calculations for the entire dataset remains computationally prohibitive, as completing high-fidelity orbit simulations across the entire range would scale to years of total computational time. The present work therefore adopts a hybrid approach in which the converged TRANSP results are used primarily to evaluate the magnitude and systematic trends of the uncertainty associated with simplified fast-ion loss models, while continuing to utilize the explicit formulation from equation~\eqref{eq:fast_ion_empirical} to anchor database regressions.

In addition to uncertainties in fast-ion losses, the definition of $P_{\mathrm{Loss}}$ itself introduces ambiguity when comparing across machines and operating regimes. Variations in equilibrium reconstruction methods, stored energy determination, radiation accounting, and beam absorption models can all influence the inferred threshold power. By restricting the present study to a single device with relatively uniform analysis methodology, these sources of hidden variability are reduced compared to multi-machine databases. The modified power accounting methodology described above was used consistently throughout the database analysis and regression studies presented in the following sections.

%% file: Section5_Results.tex
\section{Results}
\label{sec:results}

\subsection{Regression Methodology}
\label{subsec:regression}

To evaluate the predictive capability of empirical threshold scaling relations, regression analysis was performed using both the standard Martin 2008 scaling variables and extended parameter sets incorporating metrics of applied three-dimensional (3D) magnetic perturbations \cite{martin2008itpa, schmitz2019}. Following the methodology commonly used in empirical L--H threshold studies, regressions were performed in logarithmic form assuming power-law dependence between the threshold power and the selected independent variables \cite{martin2008itpa}.

For each discharge, the experimentally inferred threshold power $P_{\mathrm{Loss}}$ was compared against the predicted threshold obtained from the regression model. Unless otherwise noted, logarithmic least-squares fitting was used to determine regression coefficients of the form

\begin{equation}
P_{\mathrm{th}} = C \prod_i x_i^{\alpha_i},
\label{eq:generic_regression}
\end{equation}

where $C$ is a constant coefficient, $x_i$ represents the selected independent variables, and $\alpha_i$ are the fitted scaling exponents.

Several classes of regression models were examined. The first consisted of comparisons against the original Martin 2008 ITPA scaling without modification. Additional regressions then incorporated various perturbation metrics derived from the SURFMN analysis, including resonant magnetic field amplitude, dominant toroidal mode number, and radially localized values of $\delta B_r/B_T$ evaluated near the plasma edge.

\begin{table}[b!]
\centering
\caption{Fitted parameters and scaling exponents for unconstrained log-linear regressions executed across isolated DIII-D database subsets.}
\label{tab:regression_subsets}
\begin{tabular}{llc}
\hline
\textbf{Database Subset} & \textbf{Parameter / Metric} & \textbf{Value} \\
\hline
Nominally Axisymmetric & Coefficient $C$ & $0.12$ \\
(No RMPs)              & $\alpha_{n_e}$ & $0.71 \pm 0.49$ \\
                       & $\alpha_{B_T}$ & $0.90 \pm 0.74$ \\
                       & $\alpha_S$     & $0.83 \pm 4.36$ \\
                       & RMSE           & $81\%$ \\
\hline
Error Field Correction & Coefficient $C$ & $5.1 \times 10^{-6}$ \\
Only ($n=1,2$)         & $\alpha_{n_e}$ & $0.52 \pm 0.21$ \\
                       & $\alpha_{B_T}$ & $0.52 \pm 0.33$ \\
                       & $\alpha_S$     & $3.40 \pm 2.60$ \\
                       & RMSE           & $81\%$ \\
\hline
Full 3D Field          & Coefficient $C$ & $2.4 \times 10^{-6}$ \\
Configuration          & $\alpha_{n_e}$ & $0.52 \pm 0.20$ \\
(All RMPs)             & $\alpha_{B_T}$ & $0.51 \pm 0.28$ \\
                       & $\alpha_S$     & $3.56 \pm 2.20$ \\
                       & RMSE           & $73\%$ \\
\hline
\end{tabular}
\end{table}

Regression quality was assessed using multiple statistical metrics, including the root-mean-square (RMS) deviation in logarithmic space, the coefficient of determination ($R^2$), and residual distributions relative to the measured threshold power. Comparisons were performed for both the full database and restricted higher-confidence subsets in order to evaluate the impact of uncertainties associated with beam modulation, power accounting, and fast-ion loss corrections.

Because the present database spans a comparatively limited single-machine parameter space relative to the original ITPA multi-machine dataset, the regressions presented here are not intended to produce a new universal threshold scaling law. Instead, the primary objective is to determine whether inclusion of experimentally accessible 3D field metrics meaningfully reduces the residual variance relative to existing 0--D scaling approaches.

To reduce sensitivity to poorly constrained transitions, selected regressions were repeated after excluding discharges with large uncertainty in the absorbed power or fast-ion correction terms. In addition, regressions were performed separately for nominally axisymmetric and perturbed subsets in order to isolate the contribution of externally applied magnetic fields to the observed threshold variability.

Uncertainties associated with transition timing, absorbed power, and fast-ion losses were not propagated formally within the regression procedure. Consequently, the statistical metrics reported here should be interpreted primarily as comparative measures of relative regression performance rather than absolute uncertainty estimates for reactor extrapolation.

\subsection{Comparison with the Martin Scaling}\label{subsec:martin}

The first objective of this study was to evaluate the extent to which the standard ITPA L--H power threshold scaling reproduces modern DIII-D transition behavior across the present database. For each discharge, the measured threshold power was compared against the prediction of the Martin 2008 scaling \cite{martin2008itpa},

\begin{equation}
P_{\mathrm{th,Martin}} = 0.0488 \, n_e^{0.717} B_T^{0.803} S^{0.941},
\label{eq:martin}
\end{equation}

where $P_{\mathrm{th}}$ is expressed in MW, $n_e$ is the line-averaged density in units of $10^{20}$~m$^{-3}$, $B_T$ is the toroidal magnetic field in T, and $S$ is the plasma surface area in m$^2$.

\begin{figure}[t]
    \centering
    \includegraphics[width=0.75\textwidth]{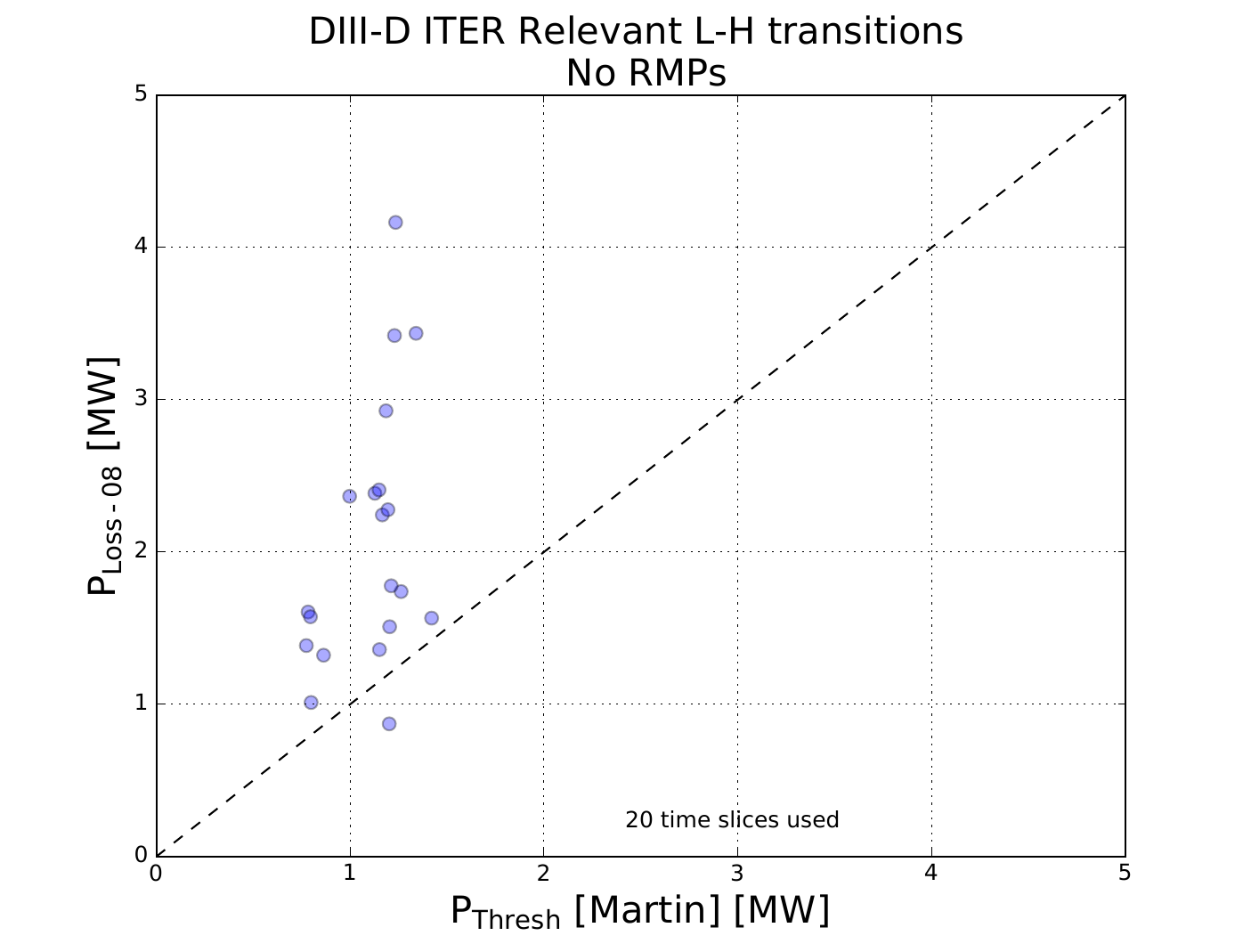}
    \caption{Measured L--H threshold power for nominally axisymmetric DIII-D discharges compared against the Martin 2008 ITPA scaling prediction. The dashed line indicates perfect 1:1 agreement. Significant scatter and systematic offset are observed even in the absence of intentionally applied resonant magnetic perturbations.}
    \label{fig:no_rmp_martin}
\end{figure}

Figure~\ref{fig:no_rmp_martin} compares the experimentally inferred threshold power against the Martin scaling prediction for the full DIII-D database. Significant scatter is observed across the dataset, including for nominally axisymmetric discharges without intentionally applied resonant magnetic perturbations. While the scaling captures the approximate global trend of increasing threshold power with density and magnetic field, the variance within the single-machine dataset remains substantial. 

To establish a clear physical baseline, an unconstrained log-linear regression was performed solely on the 20 transitions that occurred under entirely axisymmetric, non-perturbed parameters. The resulting expression exhibits a significant increase in parameter uncertainty:

\begin{equation}
P_{\mathrm{th, no\,RMP}} = 0.12 \, e^{\pm17.3} \, n_{e20}^{0.71 \pm 0.49} B_{T}^{0.90 \pm 0.74} S^{0.83 \pm 4.36}.
\label{eq:no_rmp_fit}
\end{equation}

While the calculated exponents loosely mirror the original multi-machine indices from equation~\eqref{eq:martin}, this specialized fit yields a substantial residual root-mean-square error (RMSE) of $81\%$ and projects a startup threshold power for ITER ($B_T = 5.3$~T, $S = 678$~m$^2$) at half-operational density ($n_{e20} = 0.5$) of $72$~MW, a $40\%$ inflation over standard ITPA estimates. A detailed comparison of the scaling trends is summarized in table~\ref{tab:regression_subsets}.

The observed scatter is comparable to or larger than the uncertainty typically assumed for reactor extrapolation studies based on the original ITPA regression. In particular, several discharges exhibit measured threshold powers significantly above the Martin prediction despite satisfying the ITER-relevant operating constraints used in the original multi-machine study.

To investigate whether the disagreement originates primarily from discharges with applied 3D fields, the database was subdivided into nominally axisymmetric and perturbed subsets. The axisymmetric subset shows somewhat improved agreement with the Martin scaling, although substantial variance remains. This observation suggests that the scatter cannot be attributed solely to the presence of externally applied magnetic perturbations.

\begin{figure}[t]
    \centering
    \includegraphics[width=0.78\textwidth]{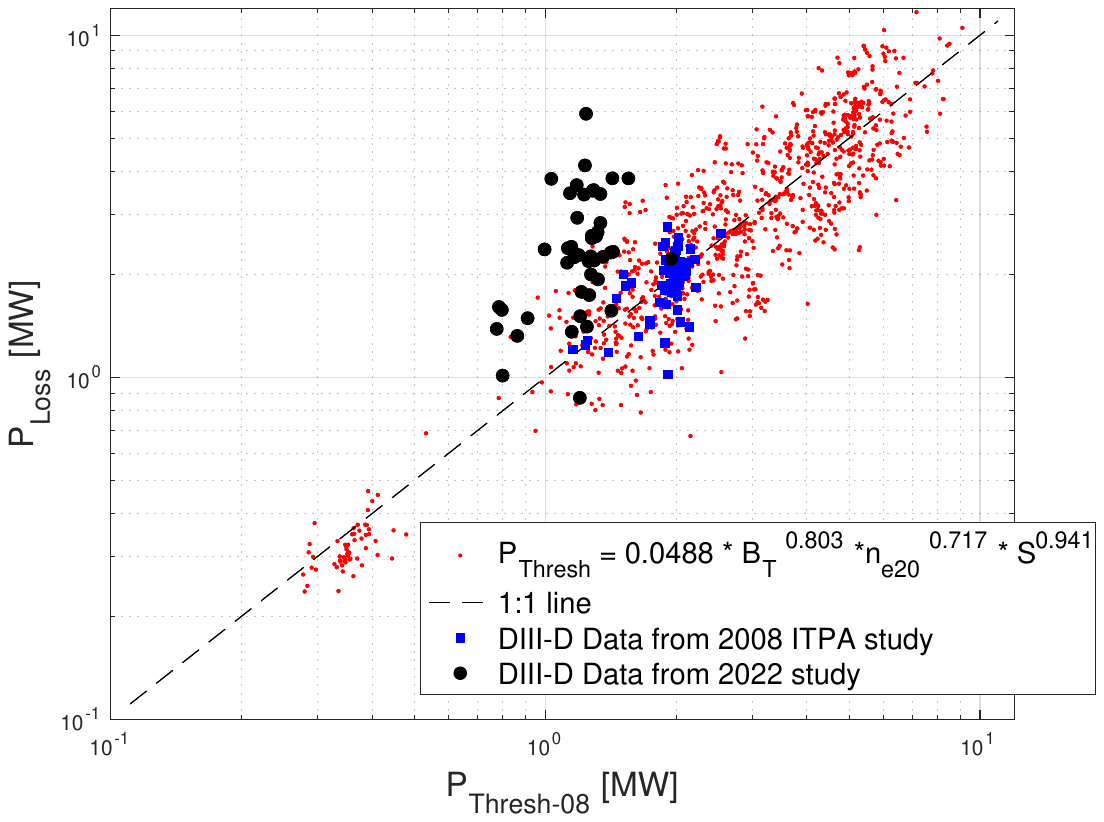}
    \caption{Comparison between the original ITPA multi-machine L--H threshold database and the updated DIII-D database constructed in the present study. The modern DIII-D transitions systematically require higher power for H-mode access than predicted by the original Martin scaling, including the subset without applied RMPs.}
    \label{fig:itpa_new_diiid}
\end{figure}

The perturbed subset generally exhibits larger deviations from the Martin prediction, particularly for discharges with strong resonant $n=3$ perturbations. However, the increase in threshold power is not universal across all perturbation configurations. Some discharges with comparatively large perturbation amplitudes exhibit only modest threshold modification, while others display threshold increases significantly larger than expected from standard scaling trends.

These results indicate that the effect of applied 3D fields on the L--H transition cannot be robustly represented by a simple multiplicative correction to existing 0--D scaling relations. Instead, the impact of RMPs appears to depend sensitively on additional edge physics parameters not explicitly included in the empirical regression.

\begin{figure}[t]
    \centering
    \includegraphics[width=1.0\textwidth]{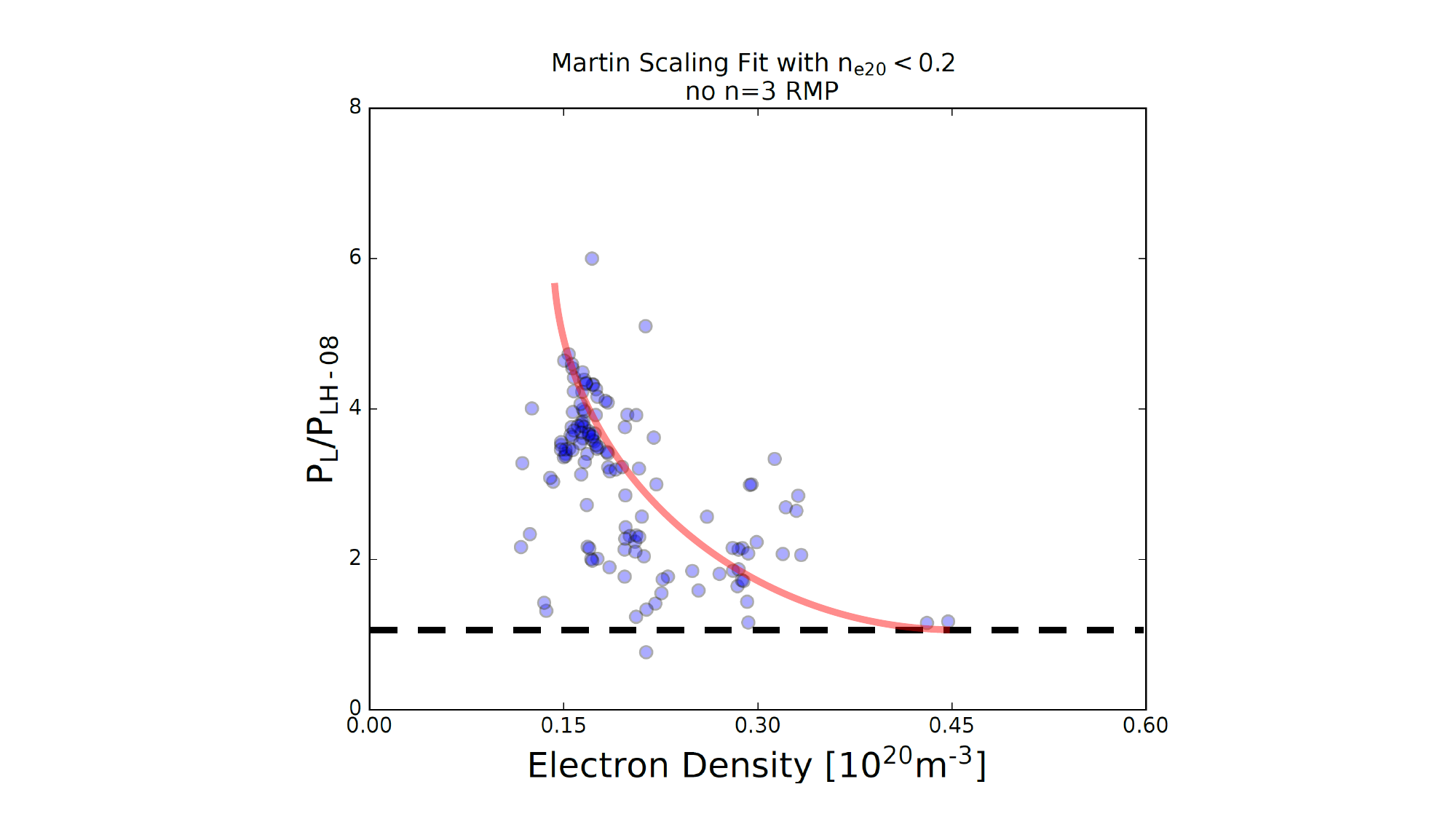}
    \caption{Ratio of the measured L–H threshold power to the Martin 2008 scaling prediction as a function of line-averaged electron density. As the electron density approaches the density branch minimum, the ratio approaches unity (dashed line). The red curve is included to guide the eye. Systematic structure in the residuals is observed, particularly near the low-density branch where the original ITPA scaling was only weakly constrained by DIII-D data.}
    \label{fig:power_ratio_density}
\end{figure}

Additional structure becomes apparent when the residuals relative to the Martin scaling are examined as a function of density. Figure~\ref{fig:power_ratio_density} shows the ratio of the measured threshold power to the Martin prediction for the subset of nominally axisymmetric DIII-D discharges. Systematic deviations are observed near the low-density branch, where several discharges require substantially larger threshold power than predicted by the original scaling.

The systematic residual structure observed near the low-density branch is particularly significant because this region was comparatively weakly constrained within the original ITPA database. The persistence of this structure within a modern single-machine dataset suggests that part of the apparent robustness of the original scaling may result from averaging over unresolved operational variability rather than from uniquely universal threshold physics.

\subsection{Dependence on Magnetic Perturbation Structure}\label{subsec:rmp}

\begin{figure}[t]
    \centering
    \includegraphics[width=0.78\textwidth]{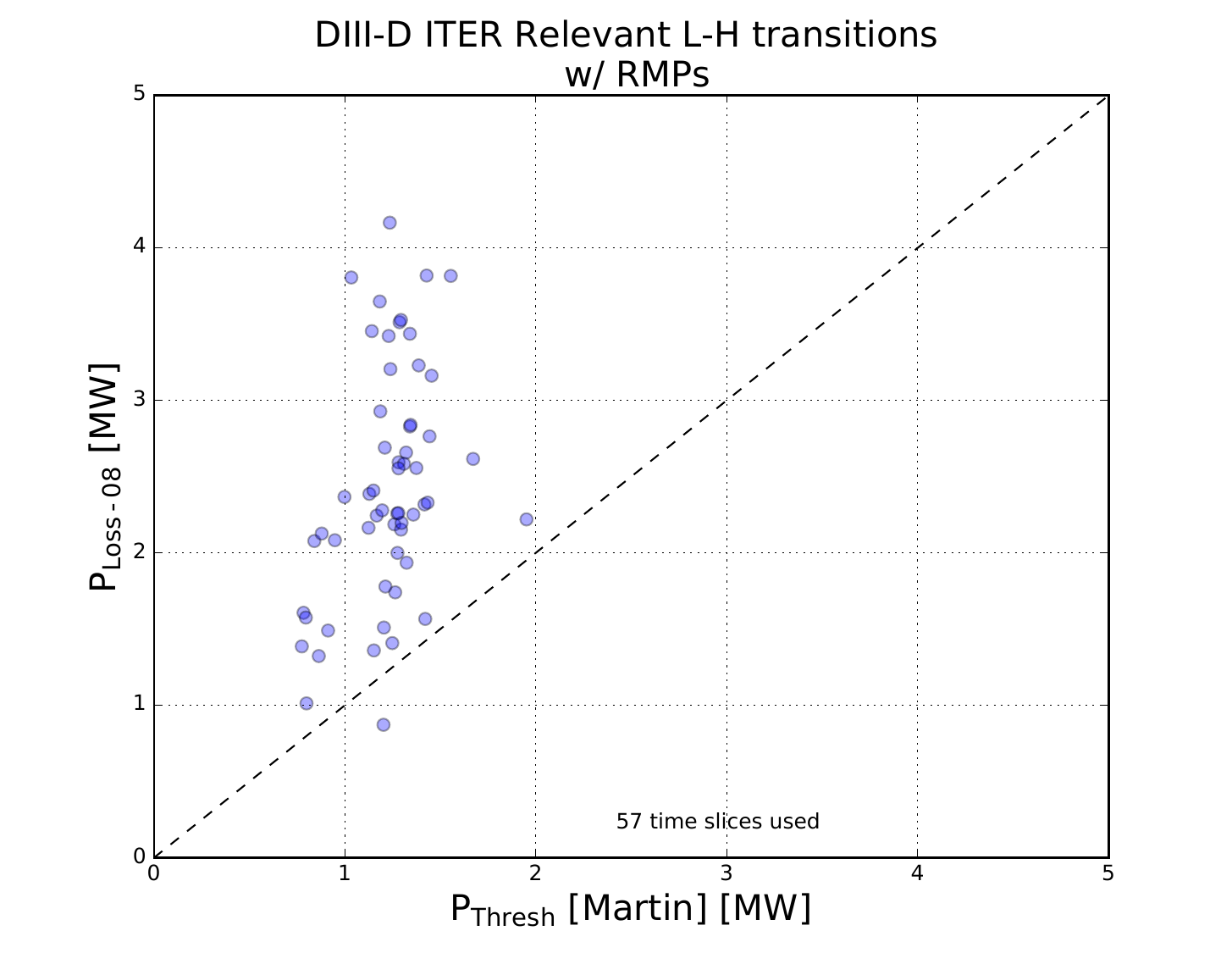}
    \caption{Full database of ITER-relevant DIII-D L--H transitions including discharges with applied resonant and non-resonant magnetic perturbations. Substantial variance relative to the Martin scaling persists across both axisymmetric and perturbed operating conditions.}
    \label{fig:full_database_rmp}
\end{figure}

To examine the relationship between the L--H threshold and the applied perturbation spectrum, several perturbation metrics derived from the SURFMN analysis were compared against the measured threshold power. These metrics included the maximum resonant field amplitude near the plasma edge, the dominant toroidal mode number, and radially localized values of $\delta B_r/B_T$ evaluated near selected rational surfaces. The full ITER-relevant database including discharges with applied resonant and non-resonant magnetic perturbations is shown in figure~\ref{fig:full_database_rmp}.

To map out the boundary changes induced by error field correction (EFC), an intermediate unconstrained regression was isolated for the 44 discharges containing solely $n=1$ and $n=2$ RMPs:

\begin{equation}
P_{\mathrm{th, EFC}} = 5.1 \times 10^{-6} \, e^{\pm10.4} \, n_{e20}^{0.52 \pm 0.21} B_{T}^{0.52 \pm 0.33} S^{3.40 \pm 2.60}.
\label{eq:efc_only_fit}
\end{equation}

When the unconstrained parameter scan is expanded to encompass the full 57-discharge database including both EFC alignment fields and deliberate $n=3$ ELM suppression perturbations, the regression yields:

\begin{equation}
P_{\mathrm{th, RMP}} = 2.4 \times 10^{-6} \, e^{\pm8.7} \, n_{e20}^{0.52 \pm 0.20} B_{T}^{0.51 \pm 0.28} S^{3.56 \pm 2.20}.
\label{eq:full_rmp_fit}
\end{equation}

Both equation~\eqref{eq:efc_only_fit} and equation~\eqref{eq:full_rmp_fit} reveal a dramatic, nonphysical migration in the effective plasma surface area exponent, expanding from its baseline near unity up to $\alpha_S \sim 3.5$. This geometric instability signals that the standard 0--D regression process is attempting to mathematically absorb non-captured localized transport physics into the available low-dimensional surface area terms.

In general, discharges with stronger resonant perturbations tended to exhibit larger threshold powers relative to the Martin prediction. This trend was most apparent for applied $n=3$ perturbations used for ELM suppression experiments. However, the correlation between threshold increase and perturbation amplitude remained relatively weak, with substantial scatter persisting even within restricted operating subsets.

The systematic residual structure observed near the low-density branch is particularly significant because this region was comparatively weakly constrained within the original ITPA database, and the standard Martin scaling fails to reflect the high and rapidly increasing $P_{\text{LH}}$ regularly observed at low densities across multiple devices. This threshold inflation is heavily coupled to the specific choice of auxiliary heating method; on DIII-D, $P_{\text{Abs}}$ can comprise varying mixes of NBI or ECH, which exhibit a massive disparity in their effectiveness for triggering the L--H transition. Because the transition is fundamentally driven by the critical ion heat flux crossing the separatrix, the poor electron--ion thermal coupling characteristic of low-density regimes means that electron-directed heating (ECH) fails to efficiently transfer energy to the main ions, thereby rapidly escalating the total required threshold power. This critical role of the separated heat channels and low-density flow-shear degradation is well-documented in recent studies by Schmitz et al.\ \cite{schmitz2019} and Yan et al.\ \cite{Yan2021IAEA}. The persistence of this residual structure within a modern single-machine dataset strongly suggests that part of the apparent robustness of the original scaling may result from averaging over such unresolved operational variability rather than from uniquely universal threshold physics.

However, the magnitude of the threshold modification associated with applied perturbations was often comparable to the intrinsic scatter already present within the nominally axisymmetric subset of the database. This observation complicates attempts to isolate a uniquely identifiable RMP contribution using only global 0--D parameters and suggests that additional local edge physics variables play an important role in determining the transition threshold.

Several factors may contribute to the observed variability. First, the externally applied vacuum perturbation does not uniquely determine the plasma response. Screening, amplification, and edge stochasticity depend strongly on plasma rotation, collisionality, pedestal structure, and edge current profiles. Second, the resonant perturbation amplitude alone does not fully characterize changes to the edge radial electric field or the associated $E\times B$ shear suppression physics thought to regulate the transition.

No single perturbation metric examined in this study was found to produce a statistically robust collapse of the threshold data across the full database. In particular, regressions including only global parameters such as maximum resonant field amplitude or toroidal mode number did not substantially reduce the variance relative to the standard Martin scaling.

These observations suggest that the impact of applied 3D fields on the L--H transition is fundamentally more complex than can be represented through simple vacuum-field metrics alone. More complete predictive models will likely require inclusion of plasma response effects and local edge transport physics. Motivated by the absence of a robust collapse using individual perturbation metrics alone, exploratory regressions were performed using combinations of global plasma parameters and resonant/non-resonant perturbation amplitudes in order to evaluate whether higher-dimensional empirical parameterizations could reduce the residual variance. To provide a concrete quantitative framework for these interactions, a comprehensive log-linear least-squares regression was executed with all global 0--D parameters and 3D field metrics unconstrained. The resulting scaling relation for the threshold power is given by:

\begin{equation}
P_{\mathrm{th}} = 5 \times 10^{-5} \, B_{T}^{0.58} n_{e20}^{0.40} S^{2.79} B_{\mathrm{R},3}^{0.11} B_{\mathrm{NR},3}^{-0.18} B_{\mathrm{R},1}^{0.22} B_{\mathrm{NR},1}^{-0.14}
\label{eq:multivariable_3d_fit}
\end{equation}

where $B_{\mathrm{R},n} = (1 + \frac{\delta B_{r}}{B_{T}} \times 10^{4})$ evaluates the normalized edge resonant harmonic component for toroidal mode $n$, and $B_{\mathrm{NR},n}$ defines the matching integrated non-resonant proxy. This unconstrained fit yields an overall RMSE of $70\%$. Notably, the positive exponents for the resonant components ($B_{\mathrm{R},3}^{0.11}$ and $B_{\mathrm{R},1}^{0.22}$) mathematically capture the systematic threshold inflation driven by edge stochasticity and flow-shear degradation, while the negative non-resonant exponents reflect the localized flow-damping or shielding trends observed in single-shot tracking. Figure~\ref{fig:itpa_new_diiid} compares the updated DIII-D database against the original ITPA threshold scaling dataset used in the Martin 2008 regression.

The unconstrained regression is shown primarily to illustrate the instability and limited physical interpretability that emerge when additional perturbation metrics are introduced into low-dimensional scaling frameworks. This expression should therefore not be interpreted as a predictive reactor scaling law, but rather as an illustration of the tendency for unconstrained regressions to absorb unresolved hidden-variable dependencies into nonphysical effective exponents.

\subsection{Role of Fast-Ion Loss Uncertainty}\label{subsec:fastion}

The comparison between empirical and TRANSP-derived fast-ion losses revealed that uncertainties in the fast-ion correction term can contribute significantly to the inferred threshold power. This effect was particularly important at lower plasma current and in discharges containing strong non-axisymmetric magnetic perturbations.

Figure~\ref{fig:transp_empirical_fastion} compares empirical fast-ion loss estimates against TRANSP calculations for a representative subset of the database. While the empirical model reproduces the approximate trend of increasing losses at lower plasma current, systematic underprediction is observed in many discharges. In several cases, the discrepancy approaches a substantial fraction of the injected neutral beam power. 

Because the fast-ion loss correction enters directly into the definition of $P_{\mathrm{Loss}}$, underestimating these losses artificially increases the inferred threshold power crossing the separatrix. This effect introduces an additional source of systematic uncertainty into empirical threshold regressions.

\begin{figure}[t]
    \centering
    \includegraphics[width=0.78\textwidth]{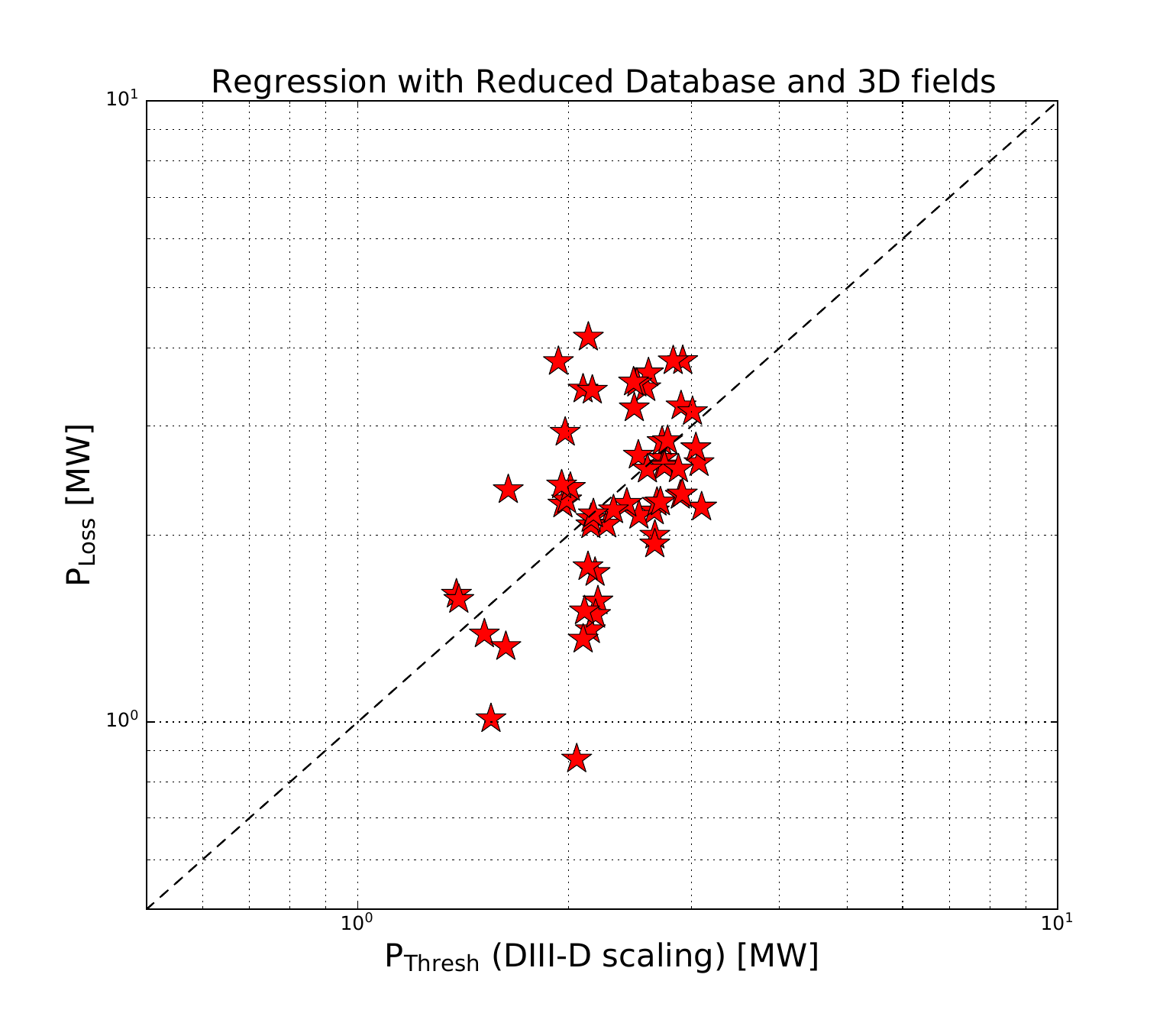}
    \caption{Measured L–H threshold power compared against the unconstrained regression model obtained from the conditioned DIII-D database including selected 3D magnetic perturbation metrics. Despite the restricted single-machine parameter space and consistent analysis methodology, substantial residual scatter remains.}
    \label{fig:diiid_regression}
\end{figure}

The impact of this uncertainty becomes particularly important when comparing discharges with and without applied 3D fields. Since resonant perturbations can enhance prompt orbit losses and modify fast-ion confinement, inaccuracies in the fast-ion correction may partially mimic or obscure genuine threshold changes associated with edge transport physics. The resulting regression performance for the conditioned DIII-D subset is shown in figure~\ref{fig:diiid_regression}.

Although the present study does not attempt to fully replace empirical fast-ion corrections with self-consistent transport modeling, the results indicate that uncertainties in beam-ion confinement represent an important hidden variable in existing L--H threshold databases. Future studies aiming to improve predictive threshold scaling may therefore require more sophisticated treatment of energetic particle losses, particularly for reactor-relevant operating scenarios with applied non-axisymmetric fields.

\subsection{Implications for Empirical Threshold Scaling}\label{subsec:implications}

Taken together, the present results demonstrate that low-dimensional empirical threshold regressions do not robustly capture the physics required to predict L--H transitions, let alone the effect of applied three-dimensional magnetic perturbations on these transitions. Even within a comparatively controlled single-machine database, substantial residual variance, unstable regression coefficients, and large extrapolation uncertainty persist despite careful filtering and consistent power accounting.

The persistence of large variance within a comparatively uniform DIII-D database implies that hidden local physics parameters play an important role in determining the transition threshold. Candidate mechanisms include edge flow shear, plasma response to externally applied perturbations, stochastic transport layer formation, divertor conditions, neutral fueling, and fast-ion confinement. Specifically, variations in the externally injected torque and the resulting edge toroidal rotation profile have been shown to drastically shift the required power threshold via momentum injection coupling \cite{mckee2009, gohil2010}. These velocity fields directly regulate the shear-induced turbulent Reynolds stress near the last closed flux surface \cite{fedorczak2012}, an interaction that is fundamentally mediated by the configuration's global magnetic geometry and cross-sectional shaping variables \cite{manz2018}, and which is directly altered by the application of resonant and non-resonant magnetic perturbations through modified edge turbulence--flow dynamics \cite{Kriete2020}. Furthermore, international cross-machine comparative studies reinforce that the precise spectral alignment of these external non-axisymmetric magnetic field components dictates whether the edge plasma screens or amplifies the fields, adding a highly localized kinetic variable that limits the utility of global 0--D regressions \cite{willensdorfer2022}.

These observations are particularly relevant for ITER and future reactor-scale devices, where externally applied magnetic perturbations are expected to be used routinely for ELM suppression and control. Since ITER operational scenarios may require application of RMP fields prior to H-mode access, uncertainties associated with threshold modification by 3D fields could influence operational margins and startup scenario development.

The present work therefore motivates development of more physics-informed approaches to L--H threshold prediction that combine empirical database methods with plasma response modeling and edge transport physics rather than relying solely on global 0--D regressions.

\subsection{Regression Instability and Reactor Extrapolation}
\label{subsec:extrapolation}

Although several regression models produced acceptable agreement within restricted regions of the DIII-D parameter space, the resulting scaling coefficients were found to be highly sensitive to the specific subset of discharges included in the fit. In particular, regressions incorporating additional 3D field metrics frequently produced large variations in the fitted exponents with comparatively small changes in the database conditioning criteria.

\begin{figure}[t]
    \centering
    \includegraphics[width=0.78\textwidth]{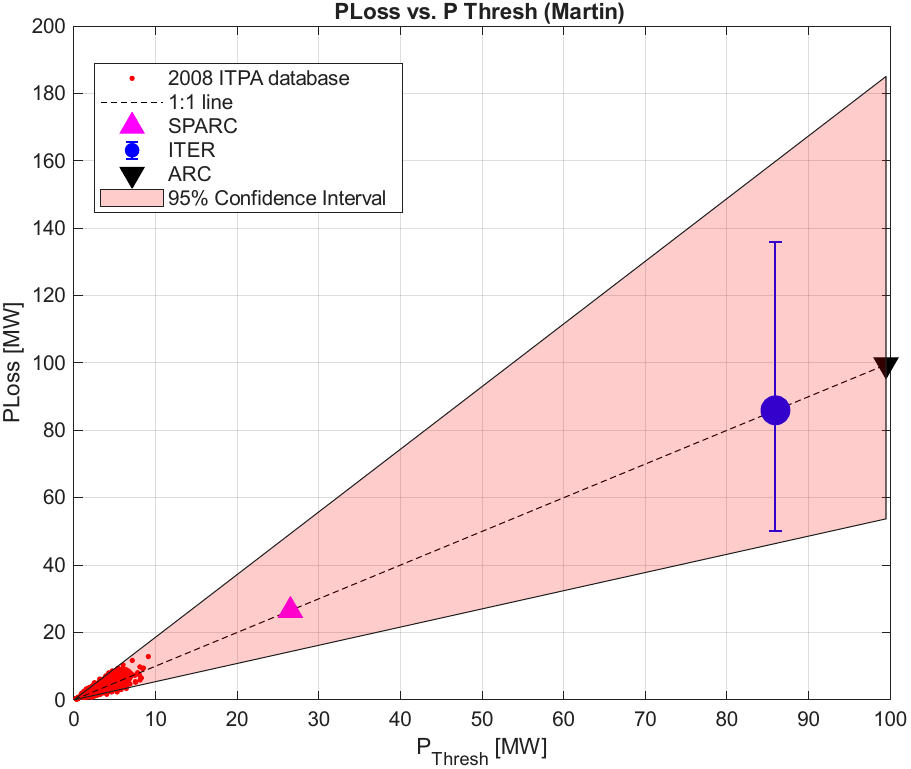}
    \caption{Extrapolation of the 2008 ITPA empirical L--H threshold scaling to reactor operating conditions. The shaded envelope defines the 95\% confidence interval, illustrating a large uncertainty spanning approximately 45--160~MW at ITER-relevant densities. Representative operating points for ITER, SPARC, and ARC are included for comparison.}    \label{fig:iter_extrapolation}
\end{figure}

This sensitivity is illustrated by the substantial variation observed in extrapolations to reactor-relevant operating conditions. Figure~\ref{fig:iter_extrapolation} shows the extrapolation obtained using the 2008 ITPA threshold scaling over a reactor-relevant density range. While the scaling reproduces the general trend of the experimental data used in its construction, the corresponding 95\% confidence interval expands rapidly outside the region populated by the original database. Representative operating points for ITER, SPARC, and ARC are included to illustrate the sensitivity of reactor-scale threshold predictions to the underlying empirical scaling.

For ITER-relevant densities near $n_{e20}\sim1$, the predicted threshold spans approximately 45--160~MW, comparable to or exceeding the anticipated auxiliary heating capability available during ITER startup scenarios. Similar uncertainty in threshold prediction would directly affect operational margins in emerging reactor concepts such as SPARC and ARC.

In several regression attempts, the fitted scaling exponents additionally evolved toward nonphysical values. In particular, unconstrained regressions sometimes produced unrealistically large dependencies on plasma surface area and magnetic geometry parameters. These trends indicate that the regression procedure attempts to absorb unresolved hidden-variable dependencies into the limited set of available global parameters.

The observed instability of the fitted coefficients suggests that low-dimensional empirical scaling approaches may not contain sufficient information to robustly parameterize the underlying transition physics once additional effects associated with plasma response, divertor geometry, and energetic particle confinement become important. While empirical regressions remain valuable engineering tools, the present results indicate that caution is required when extrapolating such models beyond the parameter space directly constrained by experimental data.

%% file: Section6_Discussion.tex
\section{Discussion}\label{sec:discussion}

The present study highlights several important limitations associated with purely empirical descriptions of the L--H power threshold in tokamaks with applied three-dimensional (3D) magnetic fields. Although the Martin 2008 scaling \cite{martin2008itpa} successfully captures broad multi-machine trends, substantial variance remains even within the comparatively uniform single-machine DIII-D database examined here. While the present database size remains modest compared to the original multi-machine ITPA compilation, the observed regression instability nevertheless persists despite substantially improved internal consistency and tighter conditioning criteria. The persistence of this scatter despite consistent equilibrium reconstruction, diagnostic methodology, and power accounting suggests that hidden local physics parameters play a major role in determining the transition threshold.

An important limitation of large multi-machine regressions is that they inherently average over discharges obtained using different diagnostics, wall materials, divertor geometries, plasma control systems, equilibrium reconstruction methodologies, and power accounting procedures. While this approach greatly expands accessible parameter space, it may simultaneously obscure machine-specific dependencies associated with local edge physics.

For example, the treatment of fast-ion losses differs substantially between devices within the original ITPA database. Some machines employed simplified algebraic estimates, while others used Monte Carlo or transport-based calculations. Similar inconsistencies exist in the treatment of radiated power corrections, wall conditions, and divertor configurations. As a result, part of the apparent success of global empirical scalings may arise from averaging over unresolved hidden-variable dependencies rather than from uniquely universal physics.

\subsection{Geometric Hardware Modifications and In-Vessel Baffle Enhancements}
\label{subsec:baffle_geometry}

A striking systematic difference between the historical DIII-D datasets included in the 2008 ITPA compilation and the modern database analyzed in this work is that the updated discharges require, on average, twice ($2\times$) the heating power to cross the separatrix and access the H-mode state. This systematic threshold inflation can be traced directly to structural variations in the in-vessel boundary and core plasma shaping parameters. The original 2008 ITPA DIII-D dataset primarily incorporated configurations from the early-to-mid 1990s (situated within the shot 80,000 range). Following the subsequent engineering installation of an extended lower divertor shelf and baffle system near shot 124,000, modern operational runs (stretching from shot 139,000 onwards) underwent permanent geometric alterations.

This hardware addition directly correlates with a $20\%$ systematic compression of the total plasma surface area, dropping from an unbaffled range of $60\text{--}67\text{~m}^2$ down to a compact envelope of $49\text{--}55\text{~m}^2$. Concurrently, the proximity to the new divertor shelf structures drove a sharp increase in the plasma boundary's triangularity. For lower single-null configurations, the average lower triangularity doubled from a baseline of $\langle\delta_{\text{lower}}\rangle = 0.33$ up to a highly shaped average of $\langle\delta_{\text{lower}}\rangle = 0.66$ within the updated dataset. 

This severe geometric modification directly influences the height of the magnetic X-point relative to the floor target tiles, an operational change known to dictate neutral particle recycling behavior and core pedestal fueling properties. Elevated X-point trajectories enhance neutral particle localized recycling dynamics near the outer midplane, effectively short-circuiting the edge flow shear and requiring a corresponding inflation in auxiliary heating power to establish a robust edge transport barrier \cite{gohil2011}.

\subsection{Stochastic Fast-Ion Transport and Multi-Harmonic Spectral Proxies}
\label{subsec:stochastic_transport}

One of the clearest observations emerging from the database analysis is that the effect of applied 3D magnetic fields cannot be represented adequately using simple vacuum perturbation metrics alone. While stronger resonant perturbations often correlate with increased threshold power, the relationship is neither universal nor monotonic. Discharges with similar vacuum perturbation amplitudes can exhibit significantly different threshold behavior.

This result is consistent with previous studies showing that the plasma response to externally applied magnetic perturbations depends strongly on local equilibrium and kinetic conditions. Screening currents, resonant field amplification, edge rotation, collisionality, and stochastic layer formation all modify the effective perturbation experienced by the edge plasma. Consequently, vacuum-field quantities such as $\delta B_r/B_T$ should be interpreted primarily as proxies for the underlying plasma response rather than complete physical descriptors.

The present results additionally suggest that energetic particle confinement represents an important hidden variable in empirical threshold studies. Comparisons with TRANSP indicate that simplified fast-ion loss models commonly used in database analyses can substantially underpredict beam-ion losses under some operating conditions, particularly at lower plasma current and in the presence of applied non-axisymmetric fields. This underprediction becomes critical under explicit RMP operating conditions; detailed orbit calculations indicate that the application of strong $n=3$ RMP ELM suppression perturbations enhances prompt orbit losses up to $8.4\%$ of the total injected neutral beam power, compared to a loss fraction of just $2.7\%$ under purely axisymmetric configurations \cite{vanzeeland2015, garciamunoz2013}.

This observation has several important implications. First, inaccuracies in the fast-ion correction directly modify the inferred value of the threshold power crossing the separatrix. Second, since applied 3D fields can themselves influence energetic particle confinement, systematic errors in fast-ion accounting may partially mimic or obscure genuine threshold modifications associated with edge transport physics. Finally, the dependence of fast-ion losses on beam geometry, plasma shaping, magnetic topology, and orbit stochasticity introduces additional machine-specific variability that is not naturally captured by standard 0--D regressions. 

This severe geometric modification directly influences the height of the magnetic X-point relative to the floor target tiles, an operational change known to strongly impact the baseline threshold power. While the precise physical mechanism behind this X-point dependence remains a subject of ongoing investigation, it is generally attributed to two competing or complementary hypotheses. First, elevated X-point trajectories can degrade divertor neutral screening, enhancing localized recycling dynamics near the outer midplane that act to mechanically dampen the edge flow shear \cite{andrew2004}. Second, changes in the X-point height and local magnetic geometry directly modify neoclassical drift orbit topologies and "X-transport" ion loss cones near the null-field region, structurally altering the intrinsic edge radial electric field profile and requiring a corresponding inflation in auxiliary heating power to establish a robust edge transport barrier \cite{Battaglia2013}.

\subsection{Analysis of Localized 0--D Scaling Exceptions}
\label{subsec:exceptions}

The fundamental inability of global, averaged 0--D scaling vectors to reliably resolve edge transport triggers is forcefully illustrated by comparing specific baseline exception pairs within the curated DIII-D database. For instance, DIII-D discharges 139979 and 139997 share virtually identical engineering parameters within standard empirical regression frameworks, tracking identical plasma surface areas ($S \sim 54\text{~m}^2$), closely matched elongations, matching safety factor constraints ($q_{95}$), and standard greenwald-normalized density lines. Consequently, the historical Martin scaling projects an identical threshold value for both shots ($P_{\mathrm{th,Martin}} \sim 1.25\text{~MW}$).

In experimental operation, however, shot 139979 crossed into a stable H-mode at a clean threshold crossing of $P_{\mathrm{Loss}} = 1.4\text{~MW}$, while its companion shot 139997 required a massive $P_{\mathrm{Loss}} = 4.2\text{~MW}$ to execute the identical transition—a factor of 3 threshold discrepancy entirely hidden from the low-dimensional empirical model. This dramatic failure underscores the sensitivity of edge transport barriers to localized, sub-centimeter modifications in magnetic geometry and target tile strike-point positions. 

Equilibrium reconstructions extracted just before the transition confirm that the low-power discharge (139979) operated with a magnetic X-point profile situated approximately $5\text{~cm}$ lower than its high-power counterpart, a variation capable of altering target neutral particle recycling distributions and pedestal fueling efficiencies across multi-megawatt scales. Furthermore, in shot 139979, the outer separatrix strike-point trajectory closely approached and potentially interacted directly with the vertical targeting face of the divertor baffle structure, modifying the local neutral recycling and fueling dynamics relative to shot 139997, where the strike point was positioned horizontally on the divertor floor target plates. This configuration difference can also alter the flux-surface averaged Reynolds stress profiles that generate the sheared $E \times B$ radial velocity flows required to trigger the transport barrier \cite{Yan2014}.

\subsection{Extrapolation Margins and Reactor Startup Boundaries}
\label{subsec:reactor_startup}

An important consequence of these findings, as mapped out by the wide residual spread in the unconstrained fits, is that uncertainties associated with reactor extrapolation may be significantly larger than commonly assumed. Existing empirical threshold scalings are often applied to ITER and future fusion pilot plant studies with comparatively modest uncertainty estimates. However, the present database demonstrates that even within a single device operating under relatively controlled conditions, threshold behavior exhibits substantial variability beyond that predicted by standard global parameters.

This issue becomes particularly relevant for ITER because operational scenarios are expected to include externally applied magnetic perturbations for ELM suppression and control. Since ITER startup procedures may require application of RMPs prior to achieving H-mode access to suppress initial transient disruptions, uncertainties in threshold modification by 3D fields could influence operational margins, auxiliary heating requirements, and startup scenario robustness.

These uncertainties are particularly important for ITER because the available auxiliary heating margin during startup is comparatively limited. Since applied 3D magnetic perturbations may be required prior to H-mode access in order to avoid the first large ELMs, uncertainty in the threshold power directly impacts operational flexibility and scenario robustness. In this regime, the bounding confidence envelope of the predicted threshold (reproduced in the unconstrained multi-machine scaling limits plotted back in Figure~9) becomes far more operationally relevant than the nominal mean scaling path itself.

The importance of robust threshold prediction extends beyond ITER and applies equally to emerging compact high-field reactor concepts such as SPARC and ARC, where operational margins between available auxiliary heating power and the L--H threshold may directly influence startup scenario flexibility \cite{creely2020, howard2026_arc}.

The present work does not imply that empirical scaling approaches are without value. On the contrary, global scaling laws remain essential for reactor design and operational planning. However, the results demonstrate that further progress in predictive threshold modeling will likely require hybrid approaches combining empirical databases with physics-based edge modeling and plasma response calculations.

Several directions for future work are suggested by this study. First, more comprehensive plasma response modeling could improve characterization of the effective perturbation experienced by the edge plasma. Second, integrated fast-ion transport calculations could reduce uncertainties associated with absorbed power and orbit losses. Third, machine-learning and reduced-order modeling approaches may provide a framework for incorporating higher-dimensional parameter dependencies beyond those captured by conventional regression techniques.

Finally, the present study demonstrates the value of carefully curated single-machine databases for isolating subtle physics effects that may be obscured within broader multi-machine regressions. While large international databases remain indispensable for reactor extrapolation, complementary single-device studies provide an important pathway toward identifying hidden variables and refining the underlying physical interpretation of empirical scaling laws.

%% file: Section7_Conclusions.tex
\section{Conclusions}\label{sec:conclusions}

A dedicated database of 192 DIII-D L--H transitions was successfully constructed and analyzed to systematically resolve the impacts of applied non-axisymmetric magnetic perturbations on the H-mode power threshold. Operating parameters were rigorously filtered down to a core conditioned baseline of 60 high-density branch, single-null discharges with favorable ion $\nabla B$ drift to ensure high-fidelity comparison across uniform multi-harmonic spectral limits. This single-device configuration framework spans a broad range of ITER-relevant conditions, successfully minimizing the hidden machine-to-machine variability that frequently compromises the interpretation of larger international compilations.

Comparisons against the standard Martin 2008 ITPA multi-machine scaling reveal substantial scatter across the entire dataset, a trend that persists even within the nominally axisymmetric, unperturbed reference discharges. While the application of edge resonant magnetic perturbations (RMPs) correlates globally with an inflation of the local threshold power crossing the separatrix, this structural behavior cannot be robustly parameterized by simple, low-dimensional vacuum-field metrics alone. No single isolated magnetic perturbation variable tracked in this study was capable of producing a statistically robust collapse of the threshold data; comprehensive multi-parameter log-linear regressions executed across all unconstrained 3D metrics left a significant residual root-mean-square error (RMSE) of 70\%.

Furthermore, the transport analysis highlights that energetic particle confinement represents a critical hidden variable within modern threshold databases. High-fidelity calculations performed with the time-dependent transport code TRANSP indicate that standard empirical algebraic expressions systematically underpredict energetic particle losses as the plasma current drops. In multiple low-current or multi-harmonic configurations, actual losses carried by charge exchange and unconfined orbits can exceed 30\% of the total injected neutral beam power—nearly doubling the rigid 16\% loss limit imposed by global empirical expressions. Underestimating these fast-ion loss channels artificially raises the inferred power flow across the boundary, distorting the empirical scaling coefficients used to model the transition.

Ultimately, these findings demonstrate that purely 0--D empirical threshold scaling laws suffer from severe predictive limitations when evaluating operating scenarios containing applied 3D non-axisymmetric fields. Unresolved local boundary variables tied directly to the self-consistent plasma response, edge stochasticity, scrape-off-layer particle recycling, and fast-ion transport profiles dominate the underlying physics of the transport barrier trigger. When projected out to baseline ITER scenarios ($B_T = 5.3$~T, $S = 678$~m$^2$), this structural low-dimensional instability inflates the calculated 95\% confidence interval to an operationally risky range of 45--160~MW at operational reactor densities. Given that the upper boundary of this empirical extrapolation more than triples the planned 50~MW auxiliary heating envelope available for ITER startup scenarios, this work strongly motivates the rapid development of non-dimensional, physics-informed threshold models that integrate empirical tracking with self-consistent edge transport and fluid response physics.

%% file: main.bbl
\begin{thebibliography}{10}

\bibitem{andrew2004}
Y.~Andrew, N.~C. Hawkes, M.~G. O'Mullane, R.~Sartori, M.~de~Baar, I.~Coffey, K.~Guenther, I.~Jenkins, A.~Korotkov, P.~Lomas, G.~F. Matthews, A.~Matilal, R.~Prentice, M.~Stamp, J.~Strachan, P.~de~Vries, and {JET EFDA Contributors}.
\newblock {JET} divertor geometry and plasma shape effects on the {L-H} transition threshold.
\newblock {\em Plasma Physics and Controlled Fusion}, 46(5A):A87--A93, 2004.

\bibitem{Battaglia2013}
D.~J. Battaglia, C.~S. Chang, S.~M. Kaye, K.~Ku, A.~Diallo, R.~Maingi, J.~E. Menard, M.~Podesta, and S.~A. Sabagh.
\newblock Dependence of the {L--H} transition on {X}-point geometry and divertor recycling on {NSTX}.
\newblock {\em Nuclear Fusion}, 53(11):113032, 2013.

\bibitem{burrell1987}
K.~H. Burrell, S.~Ejima, D.~P. Schissel, N.~H. Brooks, R.~W. Callis, T.~N. Carlstrom, A.~P. Colleraine, J.~C. DeBoo, H.~Fukumoto, R.~J. Groebner, D.~N. Hill, R.-M. Hong, N.~Hosogane, G.~L. Jackson, G.~L. Jahns, G.~Janeschitz, A.~G. Kellman, J.~Kim, L.~L. Lao, and P.~Lee.
\newblock Observation of an improved energy-confinement regime in neutral-beam-heated divertor discharges in the {DIII-D} tokamak.
\newblock {\em Physical Review Letters}, 59(13):1432--1435, 1987.

\bibitem{carlstrom1996}
T.~N. Carlstrom and R.~J. Groebner.
\newblock Study of the conditions for spontaneous {H(high)-mode} transitions in {DIII-D}.
\newblock {\em Physics of Plasmas}, 3(5):1867--1874, 1996.

\bibitem{creely2020}
A.~J. Creely, M.~J. Greenwald, S.~B. Ballinger, D.~Brunner, J.~Canik, J.~Doody, T.~F{\"{u}}l{\"{o}}p, D.~T. Garnier, R.~Granetz, T.~K. Gray, C.~Holland, N.~T. Howard, J.~W. Hughes, J.~H. Irby, V.~A. Izzo, G.~J. Kramer, A.~Q. Kuang, B.~LaBombard, Y.~Lin, and B.~Lipschultz.
\newblock Overview of the {SPARC} tokamak.
\newblock {\em Journal of Plasma Physics}, 86(5), 2020.

\bibitem{Eich2026ARC}
Thomas~H. Eich, Thomas A.~J. Body, Tom~P. Looby, Sean~B. Ballinger, Alexander~J. Creely, Jon~C. Hillesheim, Philip~C. Snyder, Nathan~T. Howard, Rebecca Masline, Michael R.~K. Wigram, and George~R. Tynan.
\newblock Power and particle exhaust for the arc fusion power plant.
\newblock {\em Journal of Plasma Physics}, 2026.
\newblock Open Access.

\bibitem{evans2004_prl}
T.~E. Evans, R.~A. Moyer, P.~R. Thomas, J.~G. Watkins, T.~H. Osborne, J.~A. Boedo, E.~J. Doyle, M.~E. Fenstermacher, K.~H. Finken, R.~J. Groebner, M.~Groth, J.~H. Harris, R.~J. La~Haye, C.~J. Lasnier, S.~Masuzaki, N.~Ohyabu, D.~G. Pretty, T.~L. Rhodes, H.~Reimerdes, and D.~L. Rudakov.
\newblock Suppression of large edge-localized modes in high-confinement {DIII-D} plasmas with a stochastic magnetic boundary.
\newblock {\em Physical Review Letters}, 92(23):235003, 2004.

\bibitem{evans2013_iter}
T.~E. Evans, D.~M. Orlov, A.~Wingen, W.~Wu, A.~Loarte, T.~A. Casper, O.~Schmitz, G.~Saibene, M.~J. Schaffer, and E.~Daly.
\newblock {3D} vacuum magnetic field modelling of the {ITER ELM} control coil during standard operating scenarios.
\newblock {\em Nuclear Fusion}, 53(9):093029, 2013.

\bibitem{fedorczak2012}
N.~Fedorczak, P.~H. Diamond, G.~Tynan, and P.~Manz.
\newblock Shear-induced {Reynolds} stress at the edge of {L-mode} tokamak plasmas.
\newblock {\em Nuclear Fusion}, 52(10):103013, 2012.

\bibitem{garciamunoz2013}
M.~{Garc{\'{i}}a-Mu{\~{n}}oz}, S.~{\"{A}}k{\"{a}}slompolo, O.~Asunta, J.~Boom, X.~Chen, I.~G.~J. Classen, R.~Dux, T.~E. Evans, S.~Fietz, R.~K. Fisher, C.~Fuchs, B.~Geiger, M.~Hoelzl, V.~Igochine, Y.~M. Jeon, J.~Kim, J.~Y. Kim, B.~Kurzan, N.~Lazanyi, and T.~Lunt.
\newblock Fast-ion redistribution and loss due to edge perturbations in the {ASDEX Upgrade}, {DIII-D} and {KSTAR} tokamaks.
\newblock {\em Nuclear Fusion}, 53(12):123008, 2013.

\bibitem{gohil2011}
P.~Gohil, T.~E. Evans, M.~E. Fenstermacher, J.~R. Ferron, T.~H. Osborne, J.~M. Park, O.~Schmitz, J.~T. Scoville, and E.~A. Unterberg.
\newblock {L-H} transition studies on {DIII-D} to determine {H-mode} access for operational scenarios in {ITER}.
\newblock {\em Nuclear Fusion}, 51(10):103020, 2011.

\bibitem{gohil2010}
P.~Gohil, T.~C. Jernigan, T.~H. Osborne, J.~T. Scoville, and E.~J. Strait.
\newblock The torque dependence of the {H-mode} power threshold in hydrogen, deuterium and helium plasmas in {DIII-D}.
\newblock {\em Nuclear Fusion}, 50(6):064011, 2010.

\bibitem{howard2026_arc}
N.~T. Howard, P.~Rodriguez-Fernandez, J.~Hall, M.~Muraca, A.~Saltzman, A.~Ho, J.~C. Hillesheim, A.~J. Creely, T.~H. Eich, T.~Body, P.~B. Snyder, and C.~Holland.
\newblock Performance and transport in the arc tokamak.
\newblock {\em Journal of Plasma Physics}, 92:E67, 2026.

\bibitem{Jackson2003}
G.~L. Jackson, M.~J. Schaffer, J.~A. Bialek, I.~N. Bogatu, M.~S. Chance, D.~H. Edgell, A.~M. Garofalo, L.~C. Johnson, I.~Joseph, G.~A. Navratil, M.~Okabayashi, H.~Reimerdes, J.~T. Scoville, and E.~J. Strait.
\newblock Initial results from the new internal magnetic field coils for resistive wall mode stabilization in the {DIII-D} tokamak.
\newblock In {\em Proceedings of the 30th EPS Conference on Controlled Fusion and Plasma Physics}, volume 27A, pages P--4.52, 2003.

\bibitem{Kriete2020}
D.~M. Kriete, L.~Schmitz, R.~S. Wilcox, T.~L. Rhodes, L.~Zeng, Z.~Yan, G.~R. McKee, R.~J. Fonck, and D.~R. Smith.
\newblock Effect of magnetic perturbations on turbulence-flow interaction at the {L-H} transition on {DIII-D}.
\newblock {\em Physics of Plasmas}, 27(6):062507, 2020.

\bibitem{Kuang2020SPARC}
A.~Q. Kuang, S.~Ballinger, D.~Brunner, J.~Canik, A.~J. Creely, T.~Gray, M.~Greenwald, J.~W. Hughes, J.~Irby, B.~LaBombard, B.~Lipschultz, J.~D. Lore, M.~L. Reinke, J.~L. Terry, M.~Umansky, D.~G. Whyte, S.~Wukitch, and the SPARC~Team.
\newblock Divertor heat flux challenge and mitigation in sparc.
\newblock {\em Journal of Plasma Physics}, 86:865860505, 2020.
\newblock Open Access.

\bibitem{lao1985}
L.~L. Lao, H.~St.~John, R.~D. Stambaugh, A.~G. Kellman, and W.~Pfeiffer.
\newblock Reconstruction of current profile parameters and plasma shapes in tokamaks.
\newblock {\em Nuclear Fusion}, 25(11):1611--1622, 1985.

\bibitem{leonard1995}
A.~W. Leonard, W.~H. Meyer, B.~Geer, D.~M. Behne, and D.~N. Hill.
\newblock {2D} tomography with bolometry in {DIII-D}.
\newblock {\em Review of Scientific Instruments}, 66(2):1201--1204, 1995.

\bibitem{ma2012}
Y.~Ma, J.~W. Hughes, A.~E. Hubbard, B.~LaBombard, R.~M. Churchill, T.~Golfinopolous, N.~Tsujii, and E.~S. Marmar.
\newblock Scaling of {H-mode} threshold power and {L--H} edge conditions with favourable ion grad-{B} drift in {Alcator C-Mod} tokamak.
\newblock {\em Nuclear Fusion}, 52(2):023010, 2012.

\bibitem{Maggi2014}
C.~F. Maggi, E.~Delabie, T.~M. Biewer, M.~Groth, N.~C. Hawkes, M.~Lehnen, E.~de la Luna, K.~McCormick, C.~Reux, F.~Rimini, E.~R. Solano, Y.~Andrew, C.~Bourdelle, V.~Bobkov, M.~Brix, G.~Calabro, A.~Czarnecka, J.~Flanagan, E.~Lerche, S.~Marsen, I.~Nunes, D.~Van Eester, M.~F. Stamp, and {JET EFDA Contributors}.
\newblock {L--H} power threshold studies in {JET} with {Be/W} and {C} wall.
\newblock {\em Nuclear Fusion}, 54(2):023007, 2014.

\bibitem{manz2018}
P.~Manz, A.~Stegmeir, B.~Schmid, T.~T. Ribeiro, G.~Birkenmeier, N.~Fedorczak, S.~Garland, K.~Hallatschek, M.~Ramisch, and B.~D. Scott.
\newblock Magnetic configuration effects on the {Reynolds} stress in the plasma edge.
\newblock {\em Physics of Plasmas}, 25(7):072508, 2018.

\bibitem{martin2008itpa}
Y.~R. Martin, T.~Takizuka, and {the ITPA Confinement Database Working Group}.
\newblock Power requirement for accessing the {H-mode} in {ITER}.
\newblock {\em Journal of Physics: Conference Series}, 123:012033, 2008.

\bibitem{mckee2009}
G.~R. McKee, P.~Gohil, D.~J. Schlossberg, J.~A. Boedo, K.~H. Burrell, J.~S. deGrassie, R.~J. Groebner, R.~A. Moyer, C.~C. Petty, T.~L. Rhodes, L.~Schmitz, M.~W. Shafer, W.~M. Solomon, M.~Umansky, G.~Wang, A.~E. White, and X.~Xu.
\newblock Dependence of the {L- to H-mode} power threshold on toroidal rotation and the link to edge turbulence dynamics.
\newblock {\em Nuclear Fusion}, 49(11):115016, 2009.

\bibitem{meneghini2015}
O.~Meneghini, S.~P. Smith, L.~L. Lao, O.~Izacard, Q.~Ren, J.~M. Park, J.~Candy, Z.~Wang, C.~J. Luna, V.~A. Izzo, B.~A. Grierson, P.~B. Snyder, C.~Holland, J.~Penna, G.~Lu, P.~Raum, A.~McCubbin, D.~M. Orlov, E.~A. Belli, N.~M. Ferraro, R.~Prater, T.~H. Osborne, A.~D. Turnbull, G.~M. Staebler, and the AToM~Team.
\newblock Integrated modeling applications for tokamak experiments with omfit.
\newblock {\em Nuclear Fusion}, 55(8):083008, 2015.

\bibitem{orlov2014}
D.~M. Orlov, R.~A. Moyer, T.~E. Evans, A.~Wingen, R.~J. Buttery, N.~M. Ferraro, B.~A. Grierson, D.~Eldon, J.~G. Watkins, and R.~Nazikian.
\newblock Comparison of the numerical modelling and experimental measurements of {DIII-D} separatrix displacements during {H-modes} with resonant magnetic perturbations.
\newblock {\em Nuclear Fusion}, 54(9):093008, 2014.

\bibitem{pankin2025}
A.~Y. Pankin, J.~Breslau, M.~Gorelenkova, R.~Andre, B.~Grierson, J.~Sachdev, M.~Goliyad, and G.~Perumpilly.
\newblock {TRANSP} integrated modeling code for interpretive and predictive analysis of tokamak plasmas.
\newblock {\em Computer Physics Communications}, 312:109611, 2025.

\bibitem{ryter2013}
F.~Ryter, S.~K. Rathgeber, L.~Barrera~Orte, M.~Bernert, G.~D. Conway, R.~Fischer, T.~Happel, B.~Kurzan, R.~M. McDermott, A.~Scarabosio, W.~Suttrop, E.~Viezzer, M.~Willensdorfer, and E.~Wolfrum.
\newblock {L--H} power threshold studies in {ASDEX Upgrade} with carbon and tungsten walls.
\newblock {\em Nuclear Fusion}, 53(11):113003, 2013.

\bibitem{ryter2013survey}
F.~Ryter, S.~K. Rathgeber, L.~Barrera~Orte, M.~Bernert, G.~D. Conway, R.~Fischer, T.~Happel, B.~Kurzan, R.~M. McDermott, A.~Scarabosio, W.~Suttrop, E.~Viezzer, M.~Willensdorfer, and E.~Wolfrum.
\newblock Survey of the {H-mode} power threshold and transition physics studies in {ASDEX Upgrade}.
\newblock {\em Nuclear Fusion}, 53(11):113003, 2013.

\bibitem{sammuli2018}
B.~S. Sammuli, J.~L. Barr, N.~W. Eidietis, K.~E.~J. Olofsson, S.~M. Flanagan, M.~Kostuk, and D.~A. Humphreys.
\newblock Toksearch: A search engine for fusion experimental data.
\newblock {\em Fusion Engineering and Design}, 129:12--15, 2018.

\bibitem{schaffer2008}
M.~J. Schaffer, J.~E. Menard, M.~P. Aldan, J.~M. Bialek, T.~E. Evans, and R.~A. Moyer.
\newblock Study of in-vessel nonaxisymmetric elm suppression coil concepts for iter.
\newblock {\em Nuclear Fusion}, 48(2):024004, 2008.

\bibitem{schmitz2019}
L.~Schmitz, D.~M. Kriete, R.~S. Wilcox, T.~L. Rhodes, L.~Zeng, Z.~Yan, G.~R. McKee, T.~E. Evans, C.~Paz-Soldan, P.~Gohil, B.~Lyons, C.~C. Petty, D.~Orlov, and A.~Marinoni.
\newblock {L-H} transition trigger physics in {ITER}-similar plasmas with applied $n=3$ magnetic perturbations.
\newblock {\em Nuclear Fusion}, 59(12):126010, 2019.

\bibitem{vanzeeland2015}
M.~A. Van~Zeeland, N.~M. Ferraro, B.~A. Grierson, W.~W. Heidbrink, G.~J. Kramer, C.~J. Lasnier, D.~C. Pace, S.~L. Allen, X.~Chen, T.~E. Evans, M.~{Garc{\'{i}}a-Mu{\~{n}}oz}, J.~M. Hanson, M.~J. Lanctot, L.~L. Lao, W.~H. Meyer, R.~A. Moyer, R.~Nazikian, D.~M. Orlov, C.~Paz-Soldan, and A.~Wingen.
\newblock Fast ion transport during applied {3D} magnetic perturbations on {DIII-D}.
\newblock {\em Nuclear Fusion}, 55(7):073028, 2015.

\bibitem{wagner1982}
F.~Wagner, G.~Becker, K.~Behringer, D.~Campbell, A.~Eberhagen, W.~Engelhardt, G.~Fussmann, O.~Gehre, J.~Gernhardt, G.~von Gierke, G.~Haas, M.~Huang, F.~Karger, M.~Keilhacker, O.~Kl{\"{u}}ber, M.~Kornherr, K.~Lackner, G.~Lisitano, G.~G. Lister, and H.~M. Mayer.
\newblock Regime of improved confinement and high beta in neutral-beam-heated divertor discharges of the {ASDEX} tokamak.
\newblock {\em Physical Review Letters}, 49(19):1408--1412, 1982.

\bibitem{willensdorfer2022}
M.~Willensdorfer, U.~Plank, D.~Brida, M.~Cavedon, G.~D. Conway, D.~A. Ryan, W.~Suttrop, R.~Buchholz, M.~Dunne, R.~Fischer, M.~Griener, J.~Hobirk, S.~Kasilov, A.~Kirk, R.~M. McDermott, T.~P{\"{u}}tterich, G.~Tardini, and Q.~Yu.
\newblock Dependence of the {L--H} power threshold on the alignment of external non-axisymmetric magnetic perturbations in {ASDEX Upgrade}.
\newblock {\em Physics of Plasmas}, 29(3):032506, 2022.

\bibitem{Yan2021IAEA}
Z.~Yan, G.~R. McKee, P.~Gohil, D.~M. Kriete, L.~Schmitz, R.~S. Wilcox, T.~L. Rhodes, L.~Zeng, and C.~C. Petty.
\newblock Turbulence flow dynamics and mode structure impacts on the {L-H} transition.
\newblock In {\em Proceedings of the 28th {IAEA} Fusion Energy Conference ({FEC} 2020)}, IAEA-CN-286, pages EX/C--2, 2021.

\bibitem{Yan2014}
Z.~Yan, G.~R. McKee, P.~Gohil, L.~Schmitz, C.~Holland, S.~R. Haskey, B.~A. Grierson, and C.~C. Petty.
\newblock Observation of the {L-H} confinement bifurcation triggered by a turbulence-driven shear flow in a tokamak plasma.
\newblock {\em Physical Review Letters}, 112(12):125002, 2014.

\bibitem{yan2019_pop}
Z.~Yan, G.~R. McKee, P.~Gohil, L.~Schmitz, S.~R. Holland, C.~Haskey, B.~A. Grierson, R.~Ke, T.~Rhodes, and C.~Petty.
\newblock Safety factor and turbulence dynamics dependence of the {L-H} power threshold on {DIII-D}.
\newblock {\em Physics of Plasmas}, 26(6):062507, 2019.

\end{thebibliography}
